\documentclass[12pt]{article}

\usepackage{cite}

\usepackage{amssymb,latexsym}
\usepackage{centernot}

\usepackage{epstopdf}

\usepackage{graphics,epsfig}

\usepackage{color}
\usepackage{xcolor}

\usepackage{amsmath,amsfonts}

\usepackage{mathrsfs}

\DeclareMathAlphabet{\mathpzc}{OT1}{pzc}{m}{it}

\usepackage{feynmp}

\let\a=\alpha \let\b=\beta \let\g=\gamma \let\d=\delta \let\e=\epsilon
\let\z=\zeta  \let\th=\theta  \let\k=\kappa
\let\l=\lambda \let\m=\mu \let\n=\nu \let\x=\xi \let\p=\pi 
\let\s=\sigma \let\t=\tau  \let\f=\phi  \let\y=\psi
 
\let\w=\omega      \let\G=\Gamma \let\D=\Delta \let\Th=\Theta \let\L=\Lambda
\let\X=\Xi  \let\S=\Sigma  \let\Y=\Psi
 
\let\la=\label  
  
\def\nn{\nonumber} \def\bd{\begin{document}} \def\ed{\end{document}}
\def\ds{\documentstyle} \let\fr=\frac \let\bl=\bigl \let\br=\bigr
\let\Br=\Bigr \let\Bl=\Bigl
\let\bm=\bibitem
\let\na=\nabla
\def\tU{{\widetilde U}}
\let\pa=\partial \let\ov=\overline
\def\ie{{\it i.e.\ }}
\newcommand{\be}{\begin{equation}}
\newcommand{\ee}{\end{equation}}
\def\ba{\begin{array}}
\def\ea{\end{array}}
\def\ft#1#2{{\textstyle{{\scriptstyle #1}\over {\scriptstyle #2}}}}
\def\fft#1#2{{#1 \over #2}}
\def\F#1#2{{ F_{#1}^{(#2)} }}
\def\cF#1#2{{ {\cal F}_{#1}^{(#2)} }}

\def\R{{\bf R}}
\def\sst#1{{\scriptscriptstyle #1}}
\def\oneone{\rlap 1\mkern4mu{\rm l}}
\def\e7{E_{7(+7)}}
\def\td{\tilde}
\def\wtd{\widetilde}
\def\im{{\rm i}}
\def\bog{Bogomol'nyi\ }
\newcommand{\ho}[1]{$\, ^{#1}$}
\newcommand{\hoch}[1]{$\, ^{#1}$}
\newcommand{\bea}{\begin{eqnarray}}
\newcommand{\eea}{\end{eqnarray}}
\newcommand{\ra}{\rightarrow}
\newcommand{\lra}{\longrightarrow}
\newcommand{\Lra}{\Leftrightarrow}
\newcommand{\ap}{\alpha^\prime}
\newcommand{\bp}{\tilde \beta^\prime}
\newcommand{\cB}{{\cal B}}
\newcommand{\cO}{{\cal O}}
\newcommand{\vecx}{\vec{x}}
\newcommand{\vecy}{\vec{y}}
\newcommand{\vecp}{\vec{p}}
\newcommand{\vecq}{\vec{q}}
\newcommand{\tr}{{\rm tr} }
\newcommand{\Tr}{{\rm Tr} }
\newcommand{\NP}{Nucl. Phys. }

\newcommand{\cL}{{\cal L}}
\newcommand{\cA}{{\cal A}}
\newcommand{\cT}{{\cal T}}
\newcommand{\cR}{{\cal R}}
\newcommand{\cD}{{\cal D}}
\newcommand{\cH}{{\cal H}}

\def\Cb{\bar{C}}

\def\sst#1{{\scriptscriptstyle #1}}
\def\0{{\sst{(0)}}}
\def\1{{\sst{(1)}}}
\def\2{{\sst{(2)}}}
\def\3{{\sst{(3)}}}
\def\4{{\sst{(4)}}}
\def\5{{\sst{(5)}}}
\def\6{{\sst{(6)}}}
\def\7{{\sst{(7)}}}
\def\8{{\sst{(8)}}}
\def\9{{\sst{(9)}}}
\def\p{{\sst{(p)}}}
\def\q{{\sst{(q)}}}
\def\ve{\varepsilon}
\def\vf{\varphi}
\def\F{\Phi}
\def\wg{\wedge}

\def\thb{\bar{\theta}}
\def\Thb{\bar{\Theta}}
\def\barp{\bar{p}}
\def\barq{\bar{q}}
\def\barc{\bar{c}}
\def\bard{\bar{d}}
\def\e{\epsilon}

\def \bi{\bibitem}
\def \la {\label}

\def \l {\lambda}
\def\foot{\footnote}
\def \tl  {{\tilde \l}}
\def \sql {{\sqrt \l}}
\def \adss {$AdS_5 \times S^5$\ }
\newcommand{\rf}[1]{(\ref{#1})}
\def \ov {\over}

\def\th{\theta}
\def\Th{\Theta}
\def\vth{\vartheta}
\def\btheta{{\bar\theta}}
\def\ttheta{{{\tilde\theta}}}
\def\bttheta{{{\bar\ttheta}}}
\def\vth{\vartheta}

\def\ra{\rightarrow}
\def\N{\nabla}
\def\F{{\cal F}}
\def\uM{\underline{M}}
\def\uA{\underline{A}}
\def\uN{\underline{N}}
\def\uP{\underline{P}}
\def\ua{\underline{a}}
\def\ub{\underline{b}}
\def\uc{\underline{c}}
\def\ud{\underline{d}}
\def\ue{\underline{e}}
\def\uf{\underline{f}}
\def\ui{\underline{i}}
\def\uj{\underline{j}}
\def\uk{\underline{k}}
\def\ul{\underline{l}}
\def\ual{\underline{\alpha}}
\def\ube{\underline{\beta}}
\def\um{\underline{m}}
\def\un{\underline{n}}
\def\up{\underline{p}}
\def\uq{\underline{q}}
\def\ur{\underline{r}}
\def\us{\underline{s}}
\def\umu{\underline{\mu}}
\def\unu{\underline{\nu}}
\def\ula{\underline{\l}}
\def\uka{\underline{\k}}
\def\usi{\underline{\s}}
\def\urh{\underline{\r}}
\def\cc{\circ}
\def\eqv{\equiv}

\def\ni{\noindent}

\def\Ep{E^{{}^{(+)}}}
\def\Em{E^{{}^{(-)}}}

\def\Mp{M^{{}^{(+)}}}
\def\Mm{M^{{}^{(-)}}}

\def \ha{{1\ov 2}}

\def\r{\rho}

\def\Y{{\rm Y}}
\def\X{{\rm X}}
\def\tY{\tilde{\rm Y}}
\def\tX{\tilde{\rm X}}
\def\dY{\dot{\rm Y}}
\def\dX{\dot{\rm X}}

\def \J {\mathcal{J}}
\def \del {\partial}

\def\dF{\dot{F}}
\def\dG{\dot{G}}
\def\df{\dot{f}}
\def \E {{\cal E}}
\def \S {{\cal S}}
\def \J {{\cal J}}

\def\ms{\mathcal{S}}
\def\mj{\mathcal{J}}
\def\soj{\fr{\ms}{\mj}}
\def \R {{\bf R}}
\def \om {\omega}
\def \bE {\bar E}
\def \x {{\cal X}}

\def \bi{\bibitem}
\def \la {\label}

\def \l {\lambda}
\def\foot{\footnote}
\def \tl  {{\tilde \l}}
\def \sql {{\sqrt \l}}
\def \adss {$AdS_5 \times S^5$\ }
\def \ov {\over}

\def \varpi {{\rm w}}

\def\thb{\bar{\theta}}
\def\Thb{\bar{\Theta}}
\def\mb{\bar{\m}}
\def\ab{\bar{\a}}
\def\zb{\bar{z}}
\def\psib{\bar{\psi}}
\def\barp{\bar{p}}
\def\barq{\bar{q}}
\def\barc{\bar{c}}
\def\bard{\bar{d}}
\def\e{\epsilon}
\def\wb{\bar{w}}
\def\lb{\bar{\l}}
\def\Jb{\bar{J}}
\def\Nb{\bar{N}}
\def\Zb{\bar{Z}}
\def\pab{\bar{\pa}}

\def\At{\tilde{A}}
\def\Bt{\tilde{B}}
\def\Ct{\tilde{C}}
\def\Dt{\tilde{D}}
\def\Et{\tilde{E}}
\def\Ft{\tilde{F}}
\def\Gt{\tilde{G}}
\def\Ht{\tilde{H}}
\def\Kt{\tilde{K}}
\def\Mt{\tilde{M}}
\def\Nt{\tilde{N}}
\def\Rt{\tilde{R}}
\def\Yt{\tilde{Y}}

\def\at{\tilde{a}}
\def\bt{\tilde{b}}
\def\ct{\tilde{c}}
\def\dt{\tilde{d}}
\def\et{\tilde{e}}
\def\ft{\tilde{f}}
\def \ztt{\tilde{\z}}
\def \zetat{\tilde{\zeta}}
\def\htil{\tilde{h}}
\def\gt{\tilde{g}}
\def\nt{\tilde{n}}
\def\mut{\tilde{\mu}}
\def\nut{\tilde{\nu}}
\def\pht{\tilde{\f}}
\def\Phit{\tilde{\Phi}}
\def\vft{\tilde{\vf}}

\def\rht{\tilde{\rho}}

\def\asth{\hat{*}}
\def\phh{\hat{\phi}}

\def\bA{{\bf A}}

\def\ola{\overleftarrow}
\def\ora{\overrightarrow}
\def\alt{\tilde{\a}}

\def\ah{\hat{a}}
\def\eh{\hat{e}}
\def\eph{\hat{\e}}
\def\ph{\hat{p}}
\def\alh{\hat{\a}}
\def\beh{\hat{\b}}
\def\gah{\hat{\g}}
\def\Fh{\hat{F}}
\def\muh{\hat{\m}}
\def\nuh{\hat{\n}}
\def\thh{\hat{\th}}
\def\rhh{\hat{\r}}
\def\dh{\hat{d}}
\def\ih{\hat{i}}
\def\jh{\hat{j}}
\def\hh{\hat{h}}
\def\nh{\hat{n}}
\def\gh{\hat{g}}
\def\kh{\hat{k}}
\def\deh{\hat{\d}}
\def\wh{\hat{w}}
\def\lah{\hat{\l}}
\def\Ah{\hat{A}}
\def\Gh{\hat{G}}
\def\Kh{\hat{K}}
\def\Nh{\hat{N}}
\def\Rh{\hat{R}}
\def\Ch{\hat{C}}
\def\Yh{\hat{Y}}
\def\Omh{\hat{\Omega}}

\def\xh{\hat{x}}

\def\ps{\rlap{\, /}\;\,p }
\def\ks{\rlap{\, /}\;\,k }

\def\gym{g_{YM}}

\def\adot{\dot{a}}
\def\bdot{\dot{b}}
\def\bpa{\bar{\pa}}

\def\pr{\prime}
\def\ssk{\medskip}
\def\clb{\color{blue}}
\def\clr{\color{red}}
\def\clg{\color{green}}
\def\clp{\color{purple}}
\def\clc{\color{cyan}}
\def\clm{\color{magenta}}
\def\cly{\color{yellow}}

\def\bfA{{\bf A}}
\def\bfB{{\bf B}}
\def\bfK{{\bf K}}
\def\bfU{{\bf U}}
\def\bfX{{\bf X}}
\def\bfY{{\bf Y}}
\def\bfZ{{\bf Z}}
\def\bfg{{\bf g}}
\def\bfn{{\bf n}}

\def\bsk{\bigskip}
\def\ssk{\medskip}

\def\Ec{{\cal E}}

\def\lar{{\langle a\rangle}}
\def\lbr{{\langle b\rangle}}
\def\ldr{{\langle d\rangle}}
\def\lcr{{\langle c\rangle}}
\def\lalr{{\langle \alpha \rangle}}
\def\lber{{\langle \beta \rangle}}
\def\lgar{{\langle \gamma \rangle}}
\def\lder{{\langle \delta \rangle}}

\begin{document}

\overfullrule=0pt
\parskip=2pt
\parindent=12pt
\headheight=0in \headsep=0in \topmargin=0in
\oddsidemargin=0in

\vspace{ -3cm}
\thispagestyle{empty}

 \vspace{0.1cm}

\setcounter{equation}{0}
\setcounter{footnote}{0}
\setcounter{section}{0}

\begin{center}

{\Large\bf Quantum-gravitational trans-Planckian radiation by a rotating black hole}

\vskip 0.8cm

\vspace{0.5cm}

A. J. Nurmagambetov$\,^{\spadesuit}$\let\thefootnote\relax\footnotetext{$^{\spadesuit}$ Also at {\it Karazin Kharkov National University, 4 Svobody Sq., Kharkov, UA 61022} \& {\it Usikov Institute for Radiophysics and Electronics, 12 Proskura St., Kharkov, UA 61085}. }
and I. Y. Park{$^\dagger$}
\\

\vspace{0.3cm}

$^{\spadesuit}$
{\it Akhiezer Institute for Theoretical Physics of
NSC KIPT,\\
1 Akademicheskaya St., Kharkov, \\ UA 61108 Ukraine \\
ajn@kipt.kharkov.ua
}

\vspace{0.3cm}
{\it {}{$^\dagger$}Department of Applied Mathematics,
Philander Smith College 
                               \\
Little Rock, AR 72223, USA \\
inyongpark05@gmail.com
}

 \vspace{.5cm}

\end{center}

 \vspace{0.1cm}

\begin{abstract}

We recently studied the energy behavior of a quantum-corrected time-dependent black hole. The system analyzed was a quantum-corrected Kerr solution that settles down to a stationary configuration as the time dependence fades out. A trans-Planckian energy scaling in the vicinity of the event horizon resulted, and we proposed the trans-Planckian radiation to be the missing link in the glowing mechanism of active galactic nuclei. The main goal of the present work is to examine the scaling and structure of the radiation by analyzing the quantum momentum density. We again observe a trans-Planckian behavior. Furthermore, the momentum density displays structures that are indicative of a disk-like configuration near the equator and a collimated outflow of matter from the poles. The bipolar outflow (disk-like structure) should be an essential part of the underlying mechanism for jets (accretion disks) of active galactic nuclei.   

\end{abstract}
\newpage





\section{Introduction}

Motivated by Firewall \cite{Almheiri:2012rt}\cite{Braunstein:2009my}, we have recently explored \cite{Park:2017dib,Nurmagambetov:2018het,Nurmagambetov:2019mih,Nurmagambetov:2019bqz,Nurmagambetov:2020} quantum-gravitational effects in the vicinity of the event horizon of a time-dependent black hole. It has been unraveled that, unlike previously thought, quantum-gravitational physics can have large effects, and it indeed does, especially in astrophysical environs. In particular, it has been shown that a time-dependent black hole generically has a trans-Planckian energy density in the Planck-scale vicinity of the event horizon. Recent works reporting Planckian or trans-Planckian energy in various related contexts include \cite{Kawai:2017txu,Kawai:2020rmt,Ho:2020cbf,Ho:2020cvn}. Such physics should lead to observable astrophysical phenomena: we have proposed that the quantum-gravitational effects should provide the sought-after missing link in the production mechanism of extreme high-energy radiation of active galactic nuclei (AGNs), such as quasars.\footnote{It is known \cite{Abraham:2007si}\cite{Dermer} that some of the ultra-high-energy cosmic rays (UHECRs) originate from active galactic nuclei. For reviews of active galatic nuclei, see, e.g., \cite{Netzer}.} In this work we continue to explore various issues and expand our previous results. The present focus is the radiation governed by quantum-gravitational generalization of the Poynting (or Umov-Poynting) vector, the quantum-gravitational momentum density vector. Interestingly, the momentum density vector reveals a structure that conforms to accretion disks and jets observed in AGNs. Based on this, we propose that the quantum-gravitational effects be essential to the three characteristic features of AGNs: extreme high-energy radiation, an accretion disk, and energetic bipolar jets.

Although the significance of such an analysis would have been compromised by the long-standing non-renormalizability of gravity, it has recently been established \cite{Park:2019amz} that the appropriately-defined physical states of a gravity theory -- which are tied with the holographic property of gravity -- are renormalizable. There are several ingredients that made the renormalizability of the physical states possible. One of them is the very identification of the physical states. Although this ingredient is not necessary for the one-loop analysis, it is essential for higher-loop extension of the renormalizability. The identification has an interesting connection with the holographic property of gravity. In turn, it has been explicitly shown that holography originates from fixing the diffeomorphism gauge symmetry. On a more technical side, the use of the ``traceless'' propagator was another crucial ingredient. In the past it was observed in \cite{Gibbons:1978ac} and \cite{Mazur:1989by} that the presence of the trace mode renders the path integral ill-defined. It has been explicitly demonstrated in a perturbative Feynman-diagrammatic analysis \cite{Park:2015ota} the mechanism by which the presence of the trace mode interferes with the 4D covariance of the effective action. It has also been shown that the renormalizability analysis can be extended to multi-loops.
With the non-renormalizability no longer hampering our efforts, several issues at one-loop have been examined, including a revisit of explicit and extensive one-loop renormalization procedure all by itself \cite{Park:2016zgt}\cite{Park:2018vci} (see \cite{Park:2019amz} for a review), the computation of the energy in the vicinity of the horizon of a time-dependent black hole \cite{Nurmagambetov:2018het}\cite{Nurmagambetov:2019mih}, and the analysis of black hole information in the quantum gravitational framework \cite{Park:2017wiw,Park:2018xtt,Park:2019lbj}.

At the technical level, the time-dependent configurations are obtained as a series-form deformation of a Kerr black hole. One crucial observation made in \cite{Nurmagambetov:2018het} was that it is the time-dependence arising with the quantum corrections that is important for the trans-Planckian scaling of the energy. In the present work we extend the quantum-deformation ansatz of \cite{Nurmagambetov:2018het} and \cite{Nurmagambetov:2019mih} in several directions: previously, the case with a vanishing classical cosmological constant and black hole charge was considered. In the present work we address the cases without such restrictions, and demonstrate that the method remains competent. 
Also, only the deformations independent of the azimuthal angle were considered before. The reason for this (and the other restrictions) was essentially simplicity: since the Kerr configuration is axisymmetric, the symmetry-preserving deformations have been preferentially considered. Lift of this restriction is another generalization discussed in the present work: by taking the case of an Einstein-scalar system, we analyze a deformed solution that does have azimuthal angle dependence.

With the robustness of the method established, we continue exploring the near-horizon region. Our previous focus was the energy density. The present focus is the radiation: the energy (and other pieces of information) of the black hole system will be radiated through the momentum density vector, a generalization of the electromagnetic Poynting vector. To isolate out the quantum-gravitational momentum density vector from the stress tensor, we conduct a 3+1 splitting of the stress tensor. A close examination of the momentum density vector unravels an intense bipolar outflow of high-energy matter as well as a disk-like structure around the equator. These features should presumably play a central role in the formation of accretion disks and production of UHECR particles and jets of AGNs.

\vspace{.3in}
The rest of the paper is organized as follows. In section 2 we review our recent sequels to set the stage for the further analysis. Here we survey the (A)dS Kerr-Newman solution. In section 2.1 we then review the technique of finding a quantum-corrected solution in a series form. Our previous work of \cite{Nurmagambetov:2019mih} covered the following case, postponing more general cases: a solution with vanishing classical part of the cosmological constant and black hole charge, $\L_0=0,\, Q=0$, respectively. In section 2.2, we demonstrate that the methodology also works in a non-vanishing classical cosmological constant and black hole charge. Afterwards, we explore another direction of generalization, the 4D deformation, by taking an Einstein-scalar system. In section 3, we analyze the quantum-gravitational momentum density. We start by conducting a 3+1 splitting of the one-loop stress tensor in section 3.1. We isolate out the quantum gravitational momentum density vector. In section 3.2 we review the geodesic both in the Eddington-Finkelstein and Boyer-Lindquist coordinates for its use in section 4. By combining the results, we analyze the $\k$-scaling of the momentum density in section 4. The $\k$-scaling of the momentum density turns out to be $\sim \fr1{\k^4}$. We note that the momentum density displays disk-like and jet-like structures. We then further contemplate the implications of the results for AGN physics. In section 5 we conclude with a summary, remarks, and further directions. Appendix A has the mode results for the 4D deformation analysis of section 2.2; Appendix B has the results of 3+1 splitting of the scalar and one-loop graviton sectors for potential future use; the Boyer-Lindquist coordinate analysis of the Kerr geometry geodesic is given in Appendix C.     

\section{Review and extension of solutions}

It has been observed in a series of our recent works that time-dependent black hole solutions display, through quantum-gravitational effects, a trans-Planckian energy scaling in the Planck-length proximity of the horizon, whereas the corresponding stationary classical solutions do not. Here we review and expand the analysis of \cite{Nurmagambetov:2019mih} to set the stage for the momentum density vector analysis in the subsequent sections. We also cover the cases postponed in \cite{Nurmagambetov:2019mih}.

Throughout, we will consider either the following system of an Einstein-Maxwell-scalar theory with a Higgs potential as a whole or certain sub-sectors thereof,
\begin{equation}
\begin{array}{lll}
&S&=\fr1{\k^2}\int d^4x \sqrt{-g}\;\Big[R-2\L\Big] +\int d^4 x \sqrt{-g}\;  \Big[c_1  R^2+c_2 R_{\m\n}R^{\m\n} +\cdots\Big] \\[1ex]
&&\hspace{-.3in}-\fr14 \int d^4x \sqrt{-g}\;F_{\m\n}F^{\m\n}  -\int d^4 x \sqrt{-g}\;\Big[|\pa_\mu \psi-iqA_\m \psi|^2 
+{\l}\Big(|\psi|^2+\fr{1}{2\l} \n^2\Big)^2   \Big].
  \la{emsactcasetwo}
\end{array}
\end{equation}
The metric, vector, and scalar field equations are, respectively,
\bea
&& R_{\m\n}-\L g_{\m\n}
-\fr{\k^2}2g_{\m\n}\Big[ {\l}\Big(|\psi|^2+\fr{1}{2\l} \n^2\Big)^2 -\fr14 F_{\a\b}F^{\a\b}  \nn\\
                       &&\hspace{1in}   +c_1  R^2+(2c_1+c_2)  \nabla^2 R  +c_2  R_{\a\b}R^{\a\b}  +\cdots \Big]  \nn\\
&&+ \k^2\Big[  { -\fr12  \left((\pa_\mu \psi-iqA_\m \psi)(\pa_\n \psi^*+iqA_\n \psi^*)+(\m \leftrightarrow \n)\right)} -\fr12 F_{\m\r}F_\n{}^\r  \nn\\
&&+2c_1  RR_{\m\n}  -(2c_1+c_2)  \nabla_\m \nabla_\n R
                             -2c_2  R_{\k_1\m\n\k_2} R^{\k_1\k_2}+c_2  \nabla^2 R_{\m\n}  +\cdots\Big]  \nn\\
                             &&=0 ,  \la{quanfe}
\eea
\[
{ \nabla^\m F_{\m\n}+iq \psi (\pa_\n+iq A_\n)\psi^*-iq \psi^* (\pa_\n-iq A_\n)\psi  +\cdots=0,}
\]
\bea
\hspace{.4in}(\nabla^\m-iqA^\m)(\nabla_\m-iqA_\m) \psi-\n^2\psi -{2}\l \psi |\psi|^2 +\cdots=0. \qquad \nn
\eea
where the $c$-coefficients are determined by one's renormalization conditions \cite{Park:2019amz}; the ellipsis stands for the term higher in $\hbar$ and/or $\k$. (The $\hbar$-dependence has been suppressed; it can be reinstated with rescaling of the $c$-coefficients by $c_1\ra \hbar c_1$, $c_2\ra \hbar c_2$.)

Note that the action \rf{emsactcasetwo} is the effective action that results from quantizing both the metric and matter fields with subsequent renormalization (see, e.g. \cite{Park:2019amz} {\clb \cite{Park:2018vci}}).\footnote{In the case of an Einstein-Maxwell system  considered in \cite{Park:2018vci} the exact values of the $c_{1,2}$ coefficients of \rf{emsactcasetwo} in the modified minimal subtraction scheme are: 
\[
c_1=\fr3{80}\,\fr{\G(2-\fr{D}2)}{(4\pi)^2},\qquad c_2=\fr{7}{40}\,\fr{\G(2-\fr{D}2)}{(4\pi)^2}.
\]
The gamma-function argument includes the number of space-time dimensions $D$.
} 
It is the leading part of the complete quantum action. The non-local terms related to anomalies, although important in various aspects of general relativity and cosmology (see, for instance, \cite{Donoghue:2017pgk} \cite{Belgacem:2017cqo} in this respect), do not qualitatively change the conclusion: non-locality is related to long-distant effects.

Throughout, we employ two coordinate systems:  Eddington-Finkelstein (EF) and Boyer-Lindquist (BL) coordinates. The EF coordinates turned out highly effective when it came to finding the quantum-corrected series solutions. Also, the metric is free from unphysical singularity at the horizon in the EF coordinates that simplifies the analysis of near horizon divergencies for other quantities, such as the energy and momentum densities. The (A)dS Kerr-Newman solution was long known in the BL coordinates. The BL coordinates will become relevant when studying the momentum density vector from the vantage point of an observer far away from the black hole. Before we get to the quantum deformations of an (A)dS Kerr-Newman solution, let us review it in each coordinate system as well as how to go back and forth between the two coordinate systems.

\vspace{.2in}
Consider an Einstein-Maxwell system. In the BL coordinates, the (A)dS Kerr-Neumann solution is given by
\cite{Hawking:1998kw} \cite{Sekiwa:2006qj}
\be
A_\m=\Bigg(\fr{Q r}{\Phi^2(r,\th)\,\Xi},0,0,-\fr{Qr a \sin^2\th}{\Phi^2(r,\th)\,\Xi} \Bigg),
\la{AmuKNBL}
\ee
\[
ds^2=\Bigg(-\fr{\D_r}{\Phi^2(r,\th)\Xi^2}+\fr{\D_\th \,a^2 \sin^2\th}{\Phi^2(r,\th)\Xi^2} \Bigg)dt^2+\fr{2a\sin^2\th}{\Phi^2(r,\th)\,\Xi^2}\left(\D_r-(r^2+a^2)\D_\th \right)dt d\vf
\]
\be
+\fr{\Phi^2(r,\th)}{\D_r} dr^2+\fr{\Phi^2(r,\th)}{\D_\th} d\th^2+\fr{\sin^2\th}{\Phi^2(r,\th)\, \Xi^2}\Bigg(-\D_r\, a^2 \sin^2\th+\D_\th (r^2+a^2)^2 \Bigg)d\vf^2,
\la{ds2KNBL}
\ee
whose matrix form is
\be
g_{\m\n}=\left(
\begin{array}{cccc}
 \frac{{{\Delta_\th} \,a^2 \sin^2\theta }-{\Delta_r}}{{\Phi^2 (r,\theta )}\Xi ^2} & 0 & 0 & \frac{a \sin^2\theta  \left({\Delta_r}-\left(r^2+a^2\right) {\Delta_\th}\right)}{ \Phi^2 (r,\theta )\,\Xi ^2} \\
 0 & \frac{\Phi^2 (r,\theta )}{{\Delta_r }} & 0 & 0 \\
 0 & 0 & \frac{\Phi^2 (r,\theta )}{{\Delta_\th }} & 0 \\
\frac{a \sin^2\theta  \left({\Delta_r}-\left(r^2+a^2\right) {\Delta_\th}\right)}{ \Phi^2 (r,\theta )\,\Xi ^2}& 0 & 0 & -\frac{\sin^2\theta  \left({\Delta_r } \,a^2 \sin^2\theta -{\Delta_\th }\left(r^2+a^2\right)^2 \right)}{ \Phi^2 (r,\theta )\,\Xi ^2} \\
\end{array}
\right)
\ee
where $a,Q$ are the rotation parameter and charge of the black hole, respectively, and
\bea
\Phi^2=r^2+a^2\cos^2\th,\quad \Xi=1+\fr{a^2}{l^2}, \quad \Delta_\th=1+\fr{a^2}{l^2}\cos^2\th ,
\la{PhiKNCCzBL}
\eea
\be
\Delta_r=\left(a^2+r^2\right)\left(1- r^2/l^2\right)-2Mr+\k^2\fr{Q^2}4.
\la{DzdefKNCCBL}
\ee
The cosmological constant $\L$ has been set to $\L=\fr{3}{l^2}$ with
\be
l^2>0\,\,\,\,\,\,(\mathrm{dS}),\qquad l^2<0\,\,\,\,\,\,(\mathrm{AdS}).
\la{ltoCC}
\ee
(Later, $\L$ is split into the classical part and quantum part: 
\be
\L=\L_0+\hbar \k^2\L_1.
\ee
Thus, it is really $\L_0$, more precisely, that is related to $l$, i.e., $\L_0=\fr{3}{l^2}$.) The standard Kerr case corresponds, of course, to $\L=0$ (or $l^2 \ra \infty$). In the EF coordinates, \cite{Lake:2015xca}
\be
A_\m=\Bigg(\fr{Q }{z\Phi^2\,\Xi},0,0,\fr{Q a \sin^2\th}{z\Phi^2\,\Xi} \Bigg),
\la{AmuKNEF}
\ee
\[
ds^2=-\fr{\Delta_z-a^2 \Delta_\th \sin^2\th}{\Phi^2 \Xi^2 } du^2-\fr{2}{z^2 \Xi } du dz-\fr{2a}{z^2 \Xi}\sin^2\th dz d\f
\]
\[
+\fr{2a\sin^2\th}{ \Phi^2\,\Xi^2}\left(\left(z^{-2}+a^2\right)\Delta_\th-\Delta_z\right)du d\f+\fr{\Phi^2}{\Delta_\th} d\th^2
\]
\be
+\fr{\sin^2\th}{\Phi^2\,\Xi^2}\left(\Delta_\th\left(z^{-2}+a^2\right)^2-\Delta_z\, a^2 \sin^2\th \right) d\f^2 
\la{ds2KNCCq}
\ee
whose matrix form is
\be
g_{\m\n}=\left(
\begin{array}{cccc}
 \frac{a^2 \Delta_\theta  \sin^2\theta -{\Delta_z}}{ \Phi^2(z,\theta )\,\Xi ^2} & -\frac{1}{z^2 \Xi } & 0 & \frac{a \sin^2\theta  \left(\left(z^{-2}+a^2\right) \Delta_\theta -\Delta_z\right)}{ \Phi^2(z,\theta )\,\Xi ^2} \\
 -\frac{1}{z^2 \Xi } & 0 & 0 & -\frac{a \sin^2\theta }{ z^2 \Xi} \\
 0 & 0 & \frac{\Phi^2(z,\theta )}{\Delta_\theta} & 0 \\
 \frac{a \sin^2\theta  \left(\left(z^{-2}+a^2\right) \Delta_\theta -\Delta_z\right)}{ \Phi^2(z,\theta )\,\Xi ^2} & -\frac{a \sin^2\theta }{ z^2 \Xi} & 0 & \frac{\sin^2\theta  \left(\Delta_\theta \left(z^{-2}+a^2\right)^2  - \Delta_z\,a^2 \sin^2\theta \right)}{ \Phi^2(z,\theta )\,\Xi ^2} \\
\end{array}
\right)
\ee
where
\be
\Phi^2=z^{-2}+a^2\cos^2\th, \quad
\Xi=1+\fr{a^2}{l^2},\quad \Delta_\th=1+\fr{a^2}{l^2}\cos^2\th ,
\la{PhiKNCCz}
\ee
\be
\Delta_z=\left(a^2+z^{-2}\right)\left(1-(z l)^{-2}\right)-\fr{2M}{z}+\k^2\fr{Q^2}4.
\la{DzdefKNCC}
\ee

\subsubsection*{Conversion between BL and EF}

One can go from the BL solution \rf{ds2KNBL} to the EF solution \rf{ds2KNCCq} by
\be
dt=du-\fr{\Xi}{\D_r} (r^2+a^2) dr,\qquad d\vf=-d\f-\fr{\Xi a}{\D_r} dr .
\la{tphi2uvarphi}
\ee
followed by $r \ra z^{-1}$ (and relabeling $\vf,t$ by $-\f,u$, respectively). The transformation that converts the EF coordinate solution \rf{ds2KNCCq} into the BL coordinate solution \rf{ds2KNBL} is\footnote{
Given that a vector field transforms according to the vector transformation law, it is puzzling that the vector fields $A_\m$ in the BL and EF coordinates, \rf{AmuKNBL} and \rf{AmuKNEF}, respectively, are related simply by $z \leftrightarrow \fr1{r}$. To see what has happened, let us consider $A_\m dx^\m$, say, in the EF coordinates,
\bea
A_\m dx^\m=A_u du+A_\f d\f
\eea
By substituting \rf{tphi2uvarphi} into this, one gets
\bea
&&A_\m dx^\m=\fr{Qr }{\Phi^2\,\Xi}dt  -\fr{Q a r\sin^2\th}{\Phi^2\,\Xi}d\vf
+\fr{Qr }{\D_r}dr.    \la{ABL}
\eea
In other words, the nonzero $r$-component, $\fr{Qr }{\D_r}dr$, appears in the BL coordinates in addition to the terms in \rf{AmuKNBL}. As one can easily check, however, the difference between \rf{AmuKNBL} and \rf{ABL} is just a gauge: their field strengths are the same.
} 
\be
du=dt+\fr{\Xi}{\D_r} (r^2+a^2) dr,\qquad d\f=-d\vf-\fr{\Xi a}{\D_r} dr.
\la{tphi2uvarphii}
\ee

\vspace{.3in}
In section 2.1 we review the technique of finding a quantum-corrected solution in a $z$-series. The analysis in \cite{Nurmagambetov:2019mih} is pushed to the higher-order modes. We then generalize the analysis in several directions in section 2.2. The generalizations are for demonstrating completeness of the method: for the study in sections 3 and 4, we consider, for technical advantages, the previous chargeless solution with zero cosmological constant, i.e., the case considered in \cite{Nurmagambetov:2019mih} (but now worked out to higher orders of $z$).

\subsection{Summary of previous works}

 Before we analyze the quantum-deformed solutions corresponding to the classical solution just listed, 
let us briefly review the works of \cite{Nurmagambetov:2018het} and \cite{Nurmagambetov:2019mih}, where the near-horizon energy measured by an infalling observer was calculated. With the cosmological constant split into the classical and one-loop parts, $\L=\L_0+\hbar \k^2\L_1$, the analysis was carried out for chargeless black holes, $Q=0$, in the background with $\L_0=0$.

A series form of a time-dependent solution was obtained by the following ansatz:
\bea
\hspace{-.5in}ds^2
&=&-\fr{F(u,z,\theta)}{z^2}du^2-\fr2{z^2}dudz+2a\Big(-  \fr{F(u,z,\theta)}{z^2}+1\Big)\sin^2\th\, dud\f -\fr{2a}{z^2}\sin^2\th dzd\f \nn\\
&&+\Phi^2(u,z,\theta) d\th^2+\Big( -\fr{a^2F(u,z,\theta)}{z^2}\sin^2\th+2a^2\sin^2\th+\Phi^2(u,z,\theta) \Big)\sin^2\th d\f^2. 
\la{Kerrans}
\nn\\
\eea
with
\begin{equation}
\begin{array}{lll}
F(u,z,\th)&=& F_0(u,\th) +F_1(u,\th) z+ F_2(u,\th)z^2+F_3(u,\th)z^3 + ...\\[1ex]
&+&\k^2  \Big[F_0^h(u,\th) +F_1^h(u,\th) z+ F_2^h(u,\th)z^2+F_3^h(u,\th)z^3 + ...\Big],\\[1ex]
\Phi(u,z,\th)&=&\dfrac{1}{z}+\Phi_0(u,\th) +\Phi_1(u,\th) z+ \Phi_2(u,\th)z^2+\Phi_3(u,\th)z^3 + ...\\[1ex]
&+& \k^2  \Big[\fr{\Phi_{-1}^h(u,\th)}{z}+\Phi_0^h(u,\th) +\Phi_1^h(u,\th) z+ \Phi_2^h(u,\th)z^2+\Phi_3^h(u,\th)z^3 + ...\Big] 
   \la{1stans}
\end{array}
\end{equation}
for the metric, and for the scalar and vector
\begin{equation}
\begin{array}{lll}
\psi(u,z,\th,\phi)&=&\psi_0(u,\th) +\psi_1(u,\th) z+ \psi_2(u,\th)z^2+\psi_3(u,\th)z^3 + ...\\[1ex]
&+&\k^2  \Big[ \psi_0^h(u,\th) +\psi_1^h(u,\th) z+ \psi_2^h(u,\th)z^2+\psi_3^h(u,\th)z^3 + ...\Big]; 
 \la{zetaser}
\end{array}
\end{equation}
\[
A_\m(u,z,\theta,\f)= (0,A_1(u,z,\theta),A_2(u,z,\theta),A_3(u,z,\theta)) 
\]
with
\begin{equation}
\begin{array}{lll}
 A_1(u,z,\theta)&=& A_{z0}(u,\th) +A_{z1}(u,\th) z+ A_{z2}(u,\th)z^2+A_{z3}(u,\th)z^3 + ... \\[1ex]
&+&\k^2  \Big[ A_{z0}^h(u,\th) +A_{z1}^h(u,\th) z+ A_{z2}^h(u,\th)z^2+A_{z3}^h(u,\th)z^3 + ...\Big],\\[1ex]
A_2(u,z,\theta)&=& A_{\th0}(u,\th) +A_{\th1}(u,\th) z+ A_{\th2}(u,\th)z^2+A_{\th3}(u,\th)z^3 + ... \\[1ex]
&+&\k^2  \Big[ A_{\th0}^h(u,\th) +A_{\th1}^h(u,\th) z+ A_{\th2}^h(u,\th)z^2+A_{\th3}^h(u,\th)z^3 + ...\Big],\\[1ex]
A_3(u,z,\theta)&=& A_{\f0}(u,\th) +A_{\f1}(u,\th) z+ A_{\f2}(u,\th)z^2+A_{\f3}(u,\th)z^3 + ...\\[1ex]
&+&\k^2  \Big[ A_{\f0}^h(u,\th) +A_{\f1}^h(u,\th) z+ A_{\f2}^h(u,\th)z^2+A_{\f3}^h(u,\th)z^3 + ...\Big],
\la{vecser}
\end{array}
\end{equation}
where the modes with superscript ``$h$'' represent the quantum modes. The $\hbar$-dependence can be made explicit by rescaling the quantum modes by 
\[
\mbox{(quantum mode)} \ra \hbar\;\mbox{(quantum mode)}.  
\]
Note that the time component of $A_\m$ is set to zero, $A_0=0$: the ansatz is adequate only for a Kerr case but not for a Kerr-Newman. 

The form of the ansatz \rf{Kerrans} should be taken to be valid only at the $z$-orders that are explicitly checked, which in our case is up to $z^4$-order. (For the scalar and vector field equations, we checked the 5th order as well.) In other words, although the ansatz \rf{Kerrans} formally takes a closed form, what is strictly valid is the series form to the order explicitly checked. Nevertheless, having a closed form ansatz -- which is more concise -- is advantageous since conciseness is huge help in machine computing. Of course, it may well also be true that the ansatz \rf{Kerrans} with \rf{1stans} and the matter fields ansatze may well remain valid, with appropriate constraints between the modes, to all orders of $z$.

Requiring that the solution settles down to a stationary configuration that includes the standard Kerr geometry (see below) as the time-dependence fades out, the first several $z$-powers of the field equations lead to a set of constraints among the modes. We have expanded them, and quote them here to point out several salient features that remain valid in the cases, to be discussed in section 2.2, of the $\L_0\neq 0,\,Q\neq 0$, and/or $\vf$-dependent deformations: for the classical modes, 
\begin{equation}
\begin{array}{lll}
&&\psi_0(u,\th)=\psi_0,\quad \psi_0 \psi_0^*=-\fr{\n^2}{2\l},\quad \\[1ex]
&& \psi_i(u,\th)=0,\,\,\,i=1, \dots, 5 ,\\[1ex]
&& A_{zi}(u,\th)=0,\quad i=0,\dots,4, \\[1ex]
&& A_{\th i}(u,\th)=0,\quad i=0,\dots,5, \\[1ex]
&& A_{\f i}(u,\th)=0,\quad i=0,\dots,5, \\[1ex]
&&  F_0(u,\th)=0,\quad  F_1(u,\theta)=0,\quad F_2(u,\theta)=1 ,\quad\\[1ex]
&& F_3(u,\theta) =-2M,\quad F_4(u,\theta) =0,\quad  F_5(u,\theta) =-a^2 \cos ^2\theta  F_3\\[1ex]
&& \Phi _{-1}(u,\theta )=1,\quad \Phi _0(u,\theta )=0,\quad  \Phi _1(u,\theta )=\fr12 a^2 \cos^2\th,\quad 
\\[1ex]
&&\Phi _2(u,\theta )=0,\quad \Phi_3(u,\theta)=-\dfrac{1}{8} a^4 \cos ^4\theta; 
\la{moderes}
\end{array}
\end{equation}
for the quantum modes,
\begin{equation}
\begin{array}{lll}
&&\pa_u \psi_0^h(u,\th)=0,\quad  \psi_0^{h*}(u,\th)= \dfrac{\nu ^2 \psi^h _0(u,\theta )}{2 \lambda  \psi _0^2},\quad \psi^{h*}_1(u,\theta )=\dfrac{\nu ^2 \psi^h_1(u,\theta )}{2 \lambda  \psi _0^2} ,  \\[1ex]
&&  \pa_u\psi^{h}_1(u,\theta ) =0,\quad \pa_u\psi^h_2(u,\theta )=0,\quad   \psi^{h*}_2(u,\theta )=\dfrac{\nu ^2 \psi^h_2(u,\theta )}{2 \lambda  \psi _0^2},\quad \\[1ex]
&&\psi^{h}_3(u,\theta)=0,\quad  \psi^{h*}_3(u,\theta)=0, \quad \psi^{h}_4(u,\theta)=0  \\[1ex]
&&\psi^{h}_5(u,\theta)=\frac{i \psi _0}{{5 \nu ^2 q}} \left(a^2 \lambda  \sin 2 \theta\,  \pa_u A^h_{\th 2}(u,\theta )+\nu ^2 q^2 A^h_{z4}(u,\theta )\right)  \\[1ex]
&&A^h_{z0}(u,\theta )=-\dfrac{i \psi^h_1(u,\theta )}{q \psi _0},\quad A_{\th0}^h(u,\th)=-\dfrac{i\pa_\th \psi^h_0{}(u,\theta )}{q \psi _0},\quad A_{\f0}^h(u,\th)=0,\quad
  \\[1ex]
&&A_{z1}^h(u,\theta)=-\dfrac{2 i \psi^h_2(u,\theta )}{q \psi _0},\quad A_{\th1}^h(u,\th)=-\dfrac{i \pa_\th\psi^h_1(u,\theta )}{q \psi _0},\quad A_{\f1}^h(u,\th)=0,
 \\[1ex]
&& A_{z2}^h(u,\theta)=
0,\quad \qquad A^h_{\f2}(u,\theta )=0,
 \\[1ex]
&&A_{\th2}^h(u,\th)=-\dfrac{i \nu ^2 q^2 \pa_\th\psi^h_2(u,\theta )}{\nu ^2 q^3 \psi _0},\quad   \\[1ex]
&& A_{z3}^h(u,\th)=0,\quad A_{\th3}^h(u,\th)=0,\quad A_{\f3}^h(u,\th)=0,\quad \\[1ex]
&& A_{\f4}^h(u,\th)=0,\quad A_{\f5}^h(u,\th)=0,  \\[1ex]
&&F_0^h(u,\th)=-\dfrac{1}{3}  \Lambda_1,\quad  F_1^h(u,\th)=-2\pa_u \Phi_{-1}^h,\quad   \\[1ex]
&&F^h_2(u,\theta)=-\dfrac{5}{3}a^2  \Lambda_1 \cos ^2\theta +2 \Phi^h_{-1}(u,\theta )-4 (\cos 2 \theta +2) \csc 2\theta \, \pa_\th\Phi^h_{-1}(u,\theta ), \\[1ex]
&&\pa_u F^h_3(u,\theta)=\frac{1}{4} \cot ^2\theta  \csc\theta  \Big[
8 (5 \cos 2 \theta +7) \sec ^3\theta\,  \pa_\th\Phi^h_{-1}(u,\theta ) \nn\\
&&+\csc \theta  \Big(6 (\cos 2 \theta +3) F_3 \pa_u\Phi^h_{-1}(u,\theta )+a^2  \Lambda_1 (-4 \cos 2 \theta +5 \cos 4 \theta +31)\Big)\Big],  \\[1ex]
&&F^h_4(u,\theta)= \frac{1}{3} \Big[24 a^2 \cot \theta  \pa_\th\Phi^h_{-1}(u,\theta )-12 a^2 \sin \theta  \cos \theta  \pa_\th\Phi^h_{-1}(u,\theta )\nn\\
&&\hspace{2in}-3 F_3 \Phi^h_0(u,\theta )+8 a^4  \Lambda_1 \cos ^4\theta \Big], \nn\\
&&\pa_\th^2\Phi^h_{-1}(u,\theta )=\dfrac{1}{4} \Big(-\cot ^2\theta  \left[3 F_3 \pa_u\Phi^h_{-1}(u,\theta )+a^2  \Lambda_1 (\cos 2 \theta +3)\right] \\[1ex]
&&\hspace{0.75in}-4 (\cos 2 \theta +3) \csc 2\theta   \,\pa_\th\Phi^h_{-1}(u,\theta )\Big) , \\[1ex]
&& \pa_u\Phi_0^h(u,\theta )=-2\Phi_{-1}^h(u,\theta )+2\cot\th \,\pa_\th\Phi_{-1}^h(u,\theta),\quad  \\[1ex]
&&\pa_\th\Phi^h_0(u,\theta )=- a^2 \sin 2\theta  \,\pa_u\Phi^h_{-1}(u,\theta ) , \\[1ex]
&& \Phi_1^h(u,\theta )=-\fr32 a^2 \cos^2\th \,\Phi_{-1}^h(u,\theta ),\quad \Phi^h_2(u,\theta)=-\dfrac{1}{2} a^2 \cos^2\theta \, \Phi^h_0(u,\theta),\quad  \\[1ex]
\end{array} 
\vspace{.05in}
\end{equation}
\begin{equation}
\hspace{-6cm}
\Phi_3^h(u,\theta)=\dfrac{7}{8} a^4 \cos^4\theta \,\Phi_{-1}^h(u,\theta).
\la{moderesh}
\end{equation}

A robust structural pattern among the mode relationships was observed in \cite{Nurmagambetov:2018het} and \cite{Nurmagambetov:2019mih}: the lowest $\hbar$- and $\k$-order terms explicitly shown in the action \rf{emsactcasetwo}, i.e., the classical action plus $R^2, R_{\m\n}R^{\m\n}$ terms, are important in determining the building blocks, such as $\Phi^h_{-1}$, of the higher modes. (The fact that the presence of $R^2,R_{\m\n}^2$ (but not the higher order terms) determine the building block modes can be seen by inspecting the structure of the field equations upon substituting the series ansatze.) One particularly novel feature is that, except $\psi_0$, the classical modes of the matter fields are removed, as can be seen from the vanishing classical mode results in eq. \rf{moderes}. In other words, if one considers purely classical field equations, the system admits nontrivial classical deformations. It is not the case once one considers the quantum-level field equations. This comes about due to the additional constraints among the classical modes, introduced by some of the leading $\hbar$-correction parts. Those additional constraints change some of the classical parts of the solution, thus rendering the classical part of the deformation vanishing.

Since we are interested in the near-horizon physics, it is convenient to introduce a new coordinate $Y$ defined by
\be
Y\equiv z-z_{EH}
\ee
and consider $Y$-expansion. In fact, we will also consider a more general expansion around an arbitrary fixed location, $z_0$. The classical location of the event horizon $z_{EH}$ is determined by $\Delta_z(z)=0$. As previously mentioned, all of the classical modes (except the irrelevant mode $\psi_0$) vanish; because of this the $Y$-series expansion can be written as
\bea
\psi(u,z,\th)&=& \tilde{\psi}_0 +\k^2\Big[\tilde{\psi}_0^h(u,\th) +  \tilde{\psi}_1^h(u,\th) Y+ \tilde{\psi}_2^h(u,\th)Y^2+\tilde{\psi}_3^h Y^3 +\cdots \Big] ,  \nn\\
\hspace{-.5in} A_1(u,z,\theta)&=& 
\k^2  \Big[ \At_{z0}^h(u,\th) +\At_{z1}^h(u,\th) Y+ \At_{z2}^h(u,\th)Y^2+\At_{z3}^h(u,\th)Y^3 + ...\Big] ,\nn\\
A_2(u,z,\theta)&=& \k^2  \Big[ \At_{\th0}^h(u,\th) +\At_{\th1}^h(u,\th) Y+ \At_{\th2}^h(u,\th)Y^2+\At_{\th3}^h(u,\th)Y^3 + ...\Big] ,\nn\\
A_3(u,z,\theta)&=& \k^2 \Big[\At_{\f6}^h(u,\th)Y^6 +\dots \Big]. 
\la{vecYsersim}
\eea
The `tilded' modes may be expressed as sums of the original modes.\footnote{In general, things become complicated very quickly in solving the field equations in the $Y$-series. In the relatively simple system of an Einstein-Maxwell considered in \cite{Nurmagambetov:2018het}, it was possible to explicitly check in the $Y$-series that the classical part of the deformation vanishes in the first several scalar modes, a result that corresponds to an all $z$-order confirmation of the vanishing of the classical part of the deformation in the $z$-series. The forms in \rf{vecYsersim} are based on the $z$-series results \rf{moderes} where the vanishing has been checked to reasonably high orders. It is not entirely clear why the $z$-series works much more effectively than the $Y$-series. It may perhaps be due to the fact that the physical states have their support on the boundary \cite{Park:2019amz}. (More on this in the Conclusion.)}

There are two noteworthy features of the quantum mode relations in \rf{moderesh}, the second of which will become quite important in later discussions. Firstly, the result, $\pa_u \psi_0^h(u,\th)=0$, implies that the building block mode $\psi_0^h(u,\th)$ -- which is the leading boundary mode -- is time-independent, but otherwise unconstrained. In particular, it does not, in general, vanish:
\be
\psi_0^h(u,\th)=\psi_0^h(\th)\neq 0.
\ee 
Therefore, there exist {\em stationary} quantum-deformed Kerr solutions. This seems to have an astrophysical implication: the ring-down phase of a black hole will not, in general, lead to the classical Kerr geometry. Instead, there is generally surviving quantum hair, although their effects may be small to observe. Secondly, various inverse powers of $\sin\th, \cos\th$ appear (often through functions such as $\cot\th$) in the constraints. For example, they appear in the right-hand sides of $F^h_2(u,\theta),\pa_\th^2\Phi^h_{-1}(u,\theta ),\pa_u\Phi_0^h(u,\theta )$ etc in \rf{moderesh}. Although the matter modes do not, to the orders obtained in \rf{moderesh}, contain any inverse power of $\sin \th$, it is expected that the higher-order modes will. This is especially the case since the different-sector modes will get widely mixed in high orders, so the metric modes, in particular, will appear in the expressions of higher matter modes. 

Due to the appearance of the inverse powers of $\sin\th$ and $\cos\th$, more care should be exercised in the small-$\th$ region, in order to ensure convergence of the series when taking various limits. What is surprising is that this subtlety may not just be a mathematical one but could be connected with the disk-like and bipolar structures of the radiation. To take a close look at these structures, it is useful to consider an expansion around an arbitrary fixed location, $z_0$,:
\bea
\psi(u,z,\th)&=& {\hat{\psi}}_0 +\k^2\Big[\hat{\psi}_0^h(u,\th) +  \hat{\psi}_1^h(u,\th) \Yh+ \tilde{\psi}_2^h(u,\th)\Yh^2+\tilde{\psi}_3^h \Yh^3 +\cdots \Big] ,  \nn\\
\hspace{-.5in} A_1(u,z,\theta)&=& 
\k^2  \Big[ \Ah_{z0}^h(u,\th) +\Ah_{z1}^h(u,\th) \Yh+ \Ah_{z2}^h(u,\th)\Yh^2+\Ah_{z3}^h(u,\th)\Yh^3 + ...\Big] ,\nn\\
A_2(u,z,\theta)&=& \k^2  \Big[ \Ah_{\th0}^h(u,\th) +\Ah_{\th1}^h(u,\th) \Yh+ \Ah_{\th2}^h(u,\th)\Yh^2+\Ah_{\th3}^h(u,\th)\Yh^3 + ...\Big] ,\nn\\
A_3(u,z,\theta)&=& \k^2  \Big[ \Ah_{\f0}^h(u,\th) +\Ah_{\f1}^h(u,\th) \Yh+ \Ah_{\f2}^h(u,\th)\Yh^2+\Ah_{\f3}^h(u,\th)\Yh^3 + ...\Big]\nn\\
\la{zzeroser}
\eea
where
\be
\hat{Y}=z-z_0.
\ee
We will come back to this series in section 4.

\vspace{.1in}

In the analysis in \cite{Nurmagambetov:2018het}, the cosmological constant $\L$ was set to $\L= \L_0+\hbar \k^2 \L_1 $ with {\em vanishing $\L_0$} to prevent occurrence of the undesirable feature noted for the Einstein-scalar system with a nonzero scalar mass. (Actually, the undesirable feature is not present for a massless scalar with a Higgs-type potential, such as the system of \cite{Nurmagambetov:2019mih}. Nevertheless, the $\L_0=0$ case is simpler and the analysis in \cite{Nurmagambetov:2019mih} was carried out by maintaining the condition $\L_0=0$.) The form of the vector field ansatz does not cover the charged black hole case since for that case one should have $A_0\neq 0$. Another restriction was that the azimuthal angle-dependence of the deformation was not considered. We now turn to lift of these conditions.

\subsection{Lift of restrictions}

In \cite{Nurmagambetov:2019mih} we analyzed, as just reviewed, quantum correction of a chargeless black holes with a vanishing classical cosmological constant, i.e., a black hole with $Q=0,\L_0=0$. Due to the vanishing cosmological constant, the time-dependent solution settles down to
a usual Kerr as opposed to a dS/AdS Kerr. To demonstrate completeness of the method, we repeat here the analysis of the cases where the black hole is charged and/or lies in a background with $\L_0\neq 0$. A series-form ansatz can be written down and substituted into the field equations. The classical part of the cosmological constant appears in inverse powers in some of the terms (although we do not explicitly record the mode relationships). In other words, the $\L_0=0$ and $\L_0\neq 0$ solutions belong to distinct branches: extending the analysis to the case of a nonvanishing classical cosmological constant is thus a meaningful exercise in this sense as well. Similarly, the charged case exhibits its own peculiarities as we examine below. Lastly, we take up another case not covered in the previous works, the generalization to the azimuthal angle-dependent deformation.

\subsubsection*{${\mathbf \L_0\neq 0}$ or ${\mathbf Q\neq 0}$}

It turns out that once one keeps $\L_0$, which makes the classical solution substantially more complicated, the computation become much more memory-demanding (even when the charge is set to zero). We thus consider the two cases, $\L_0\neq 0$ and $Q\neq 0$, separately. Also, instead of explicitly presenting the mode results here, we are content to note some of the salient features of the analyses. For the case of a nonzero cosmological constant, one complication is that since the $\D_z$ in \rf{DzdefKNCC} has lower $z$-power terms as compared to the $l\ra \infty$, this should be reflected in the form of the $\D_z$ ansatz. (We will come back to this below when we describe the 4D deformations.) We checked the case of an Einstein-scalar system without the vector field; the mode relations are obtained similarly as before.

As for the charged case, the vector field ansatz requires choosing a different gauge, as anticipated in \cite{Nurmagambetov:2019mih}. With the same forms of the ansatze for the metric and scalar as before, the ansatz for the vector field is modified to 
\bea
A_\m(u,z,\theta,\f)&=&(A_0(u,z,\theta), A_1(u,z,\theta),A_2(u,z,\theta),A_3(u,z,\theta)) \nn\\
A_0(u,z,\theta)
               &=& \fr{Q }{z(z^{-2}+a^2\cos^2\th)}. 
\eea
In other words, the $A_0$ component of the vector field is taken as the background of the Kerr geometry {\em without} the fluctuation; the fluctuation can be viewed as having been gauged away. With this arrangement the analysis can be repeated, and the mode relationships similar to the previous ones are obtained. Here we will just note some peculiarities of the results without explicitly presenting the mode relationships.

One of the lower-level mode relationships is,
\bea
\pa_u\psi _1(u,\theta )=i q Q \psi _0
\eea
where $q$, which appears in the action \rf{emsactcasetwo}, denotes the charge of the scalar field. This implies that $\psi _1(u,\theta )$ amplifies without bound as the time $u$ increases. Such a solution cannot be viewed as a small deformation of the original Kerr geometry, although it may have some different uses. One can consider setting $q=0$ to avoid such a behavior. Since some of the relationships contain $\fr1{q}$, it is necessary to run the analysis from the beginning after setting $q$ to $q=0$. The resulting outcomes are much less constraining since many of the relationships before setting $q=0$ contain the factor $q$ in front; those expressions no longer yield any constraints. For this reason one must go to higher orders once one sets $q=0$.

\subsubsection*{4D perturbation}

The deformations are taken to be independent of the azimuthal angle $\vf$ in the cases considered thus far. The analysis can be straightforwardly extended to $\vf$-dependent deformations, and we do that for the $\L_0\neq 0$ case. As mentioned before, with $\L_0\neq 0$ the computation becomes substantially more memory-demanding: we take the Einstein-scalar sub-sector of the system \rf{emsactcasetwo} instead. This generalization should also be useful for the purpose of studying the configurations with more general boundary conditions, although we will not pursue that task in this work. We take the following ansatz:
\bea
\Delta_z(u,z,\theta,\phi)
&=& \fr{\D_{-4}(u,\th,\f)}{z^{4}}+\fr{\D_{-3}(u,\th,\f)}{z^{3}}+\fr{\D_{-2}(u,\th,\f)}{z^{2}}+\cdots \nn\\
&&+\k^2  \Big[\fr{\D^h_{-4}(u,\th,\f)}{z^{4}}+\fr{\D^h_{-3}(u,\th,\f)}{z^{3}}+\fr{\D^h_{-2}(u,\th,\f)}{z^{2}}+\cdots\Big]  \la{dz}\nn\\
\eea
\begin{equation}
\begin{array}{lll}
\Phi(u,z,\th,\f)&=&\dfrac{\Phi_{-1}(u,\th,\f)}{z}+\Phi_0(u,\th,\f) +\Phi_1(u,\th,\f) z+ \Phi_2(u,\th,\f)z^2 + ...\\[1ex]
&+& \k^2  \Big[\fr{\Phi_{-1}^h(u,\th,\f)}{z}+\Phi_0^h(u,\th,\f) +\Phi_1^h(u,\th,\f) z+ \Phi_2^h(u,\th,\f)z^2 + ...\Big] 
   \la{1stans4D}
\end{array}
\end{equation}
\begin{equation}
\begin{array}{lll}
\psi(u,z,\th,\phi)&=&\psi_0(u,\th,\f) +\psi_1(u,\th,\f) z+ \psi_2(u,\th,\f)z^2 + ...\\[1ex]
&+&\k^2  \Big[ \psi_0^h(u,\th,\f) +\psi_1^h(u,\th,\f) z+ \psi_2^h(u,\th,\f)z^2 + ...\Big]. 
 \la{zetaser4D}
\end{array}
\end{equation}
Note that the ansatz for $\D_z$ starts with the $\fr1{z^4}$-term to reflect the form of the classical geometry when the classical cosmological constant is present. Repeating the analysis, one gets the mode relations whose explicit results can be found in Appendix A. One important feature, which is shared by the other cases, is that various inverse powers of $\sin\th$ appear. (To the orders examined, no inverse power of $\cos\th$ appears. It will be interesting to see whether or not they appear in higher orders. More on this in the Conclusion.)

\section{Momentum density and geodesic 4-velocity}

The main goal of the present work, which we take up in the next section, is to study the radiation emitted by a time-dependent black hole. To that end, one of the crucial steps is isolating out the momentum density vector from the stress tensor. The task can be conveniently carried out in the `covariant 3+1 splitting.' As with the energy density, the  momentum density vector consists of two parts: the classical part and quantum corrections. In section 3.1 we review the splitting, which then yields the momentum density vector, eq. \rf{Qadef}. Since we were interested in the energy density -- which is a scalar quantity -- in our previous works, the geodesic in the EF coordinates was sufficient. This, however, is not the case for the present work since the momentum density is a vector quantity; the coordinate system employed matters. It is thus desirable to have the geodesic in the coordinate system adapted to the observer far away from the black hole: we review the geodesic in the BL coordinate system.

\subsection{Quantum-gravitational momentum density}

Given a metric $g_{\m\n}$ and a four-velocity $u^\m$ associated with a congruence of curves, one can construct projection operators onto the four-velocity direction and the directions normal to the velocity, respectively. The decomposition of the stress tensor, from which the momentum density as well as other useful quantities are obtained, can be achieved by applying an appropriate combination of the projections operators. The first step is to split the metric as
\be
g_{\m\n}=h_{\m\n}-u_\m u_\n
\la{gab}
\ee
where the $h_{\m\n}$ denotes the induced metric of the 3D hypersurface; the velocity vector $u_\m$  is normalized as $u_\m u^\m=-1.$ It is orthogonal to the hypersurface spanned by $h_{\m\n}$, $u^\m h_{\m\n}=0$. (Later we will take $u^\m$ to be $u^\m=U^\m$, the geodesic four-velocity vector of the observer.) Let us define
\be
V_{\m\n}\equiv -u_\m u_\n
\ee
where ${V_\n}^\m$ is the projection operator onto the velocity. The other projection operator is ${h_\m}^\n$, where the raising of the indices is carried out by $g^{\m\n}$. These operators satisfy
\be
{V_\m}^\n {V_\n}^\r={V_\m}^\r,\quad {h_\m}^\n {h_\n}^\r={h_\m}^\r,\quad  
{V_\m}^\n {h_\n}^\r=0.
\la{project}
\ee
The 3+1 splitting of a tensor $W_{\a\b\g\dots}$ can be performed by expanding the right-hand side of the following identity,
\be
W_{\a\b\g\dots}=\left({h_\a}^\d+{V_\a}^\d\right)\left({h_\b}^\e+{V_\b}^\e\right)\dots \left({h_\g}^\l+{V_\g}^\l\right)W_{\d\e\l\dots}
\la{tensorsplit}
\ee
Applying the identity to the stress tensor $T_{\m\n}$, one gets
\be
T_{\m\n}=\left(T_{\r\s} u^\r u^\s \right) u_\m u_\n+\left(P_\m u_\n+u_\m P_\n \right)+{h_\m}^\r{h_\n}^\s T_{\r\s},
\la{Tabsplitg}
\ee
where we have introduced the momentum density vector,
\be
P_\m\equiv -{h_\m}^\r T_{\r\s} u^\s .
\la{Qadef}
\ee
For the present system, the stress tensor is given by
\footnote{Taking into account the quantum part of the stress tensor means adding the vev of the quantum fields stress tensor over to its classical part (see, e.g., \cite{Park:2019amz} for details).} 
\bea
{ T_{\m\n}} &=&- \fr2{\k^2}\L g_{\m\n}+g_{\m\n}\Big[-|\pa_\r \psi-iqA_\r \psi|^2 -{\l}\Big(|\psi|^2+\fr{1}{2\l} \n^2\Big)^2 -\fr14 F_{\r\s}^2  \nn\\
 &&\hspace{.6in} + \Big(c_1R^2-(4c_1+c_2)\nabla^2 R  +c_2 R_{\r\s}R^{\r\s}\Big)+\cdots \Big] 
\la{quanset}
\eea
\[\hspace{-.2in} +\Big[ { \left((\pa_\mu \psi-iqA_\m \psi)(\pa_\n \psi^*+iqA_\n \psi^*)+(\m \leftrightarrow \n)\right)}+  F_{\m\r}F_\n{}^\r
\]
\[
\hspace{.3in}-2 \Big(2c_1 RR_{\m\n}  -(2c_1+c_2) \nabla_\m \nabla_\n R
-2c_2 R_{\k_1\m\n\k_2} R^{\k_1\k_2}+c_2\nabla^2 R_{\m\n}\Big)  +\cdots \Big].  
\]
For our purpose this is the expression in eq. \rf{Qadef} to which we will return for further analysis in section 4. However, for its clearer physical meaning, in particular, its meaning as a generalization of the electromagnetic Poynting vector, one may further split each sector of $T_{\m\n}$. Let us illustrate the procedure with the Maxwell's sector; one of course gets the usual Poynting vector in terms of the electric and magnetic fields in the curved background. The corresponding analyses for the other sectors can be found in Appendix B.

Applying the identity \rf{tensorsplit} to the Maxwell sector stress-tensor
\be
T_{\m\n}^{(e.m.)}=F_{\m\r}{F_\n}^\r-\fr14 g_{\m\n} F_{\r\s}F^{\r\s}
\la{TEM}
\ee
one gets
\be
T_{\m\n}^{(e.m.)}=\left(T_{\r\s}^{(e.m.)} u^\r u^\s \right) u_\m u_\n+\left(P_\m^{(e.m.)} u_\n+u_\m P_\n^{(e.m.)} \right)+{h_\m}^\r{h_\n}^\s T_{\r\s}^{(e.m.)},
\la{Tabsplitem}
\ee
where 
\be
P_\m^{(e.m.)}=-{h_\m}^\r T_{\r\s}^{(e.m.)} u^\s .
\la{Qadefem}
\ee
The Maxwell field strength in terms of $E_\m$ and $H^\r$ fields (see, e.g.,\cite{Tsagas:2004kv},\cite{Ellis:1998ct}) is:
\be
F_{\m\n}=\left(u_\m E_\n-u_\n E_\m \right)+\e_{\m\n\r} H^\r\equiv 2u_{[\m} E_{\n]}+\e_{\m\n\r} H^\r
\la{FinEH}
\ee
where $-u^\r F_{\r\m}\equiv E_\m$ and ${h_\m}^\r {h_\n}^\s F_{\r\s}\equiv \e_{\m\n\r}H^\r$. It is straightforward to show
\be
T_{\m\n}^{(e.m.)}=\fr12 \left(E^2+H^2 \right)u_\m u_\n+2P^{(e.m.)}_{(\m}u_{\n)}+\fr16 \left(E^2+H^2 \right)h_{\m\n}+{\cal P}_{(\m\n)}^{(e.m.)},
\la{TEMsplit}
\ee
where ${\cal P}_{(\m\n)}$ is a symmetric, trace-free tensor
\be
{\cal P}_{(\m\n)}=\fr13 \left(E^2+H^2 \right)h_{\m\n}-E_\m E_\n-H_\m H_\n .
\la{Pabdef}
\ee

\subsection{Geodesics in EF and BL}

To compute the decomposed components of the stress tensor on the right-hand side of eq. \rf{Tabsplitg}, one needs the four-velocity vector, $U_\m$. As noted in our previous works and reviewed in section 2, the time-dependent pieces of the classical part of the quantum-level solution become constrained to vanish: the time-dependent part of the solution comes only from the quantum correction pieces. This implies that the classical part of the stress-energy is that of a Kerr geometry. Since the stress-energy tensor vanishes for a Kerr geometry, one can use the geodesic analysis of the {\em Kerr spacetime} to compute the leading quantum-gravitational correction of the energy. (This was noted in our previous works. The same is true for the momentum density. As a matter of fact, the leading behavior of the momentum density $P_\m$ is $P_\m\sim - U_\m\, \r$, as shown in \rf{Qadeflead} below, so the result for the energy density $\r$ can be borrowed from our previous work, \cite{Nurmagambetov:2019mih}.) In this subsection, we review the geodesic four-velocity both in the EF and especially BL coordinates.

The metric admits two integrals, the energy and angular momentum:
\be
p_t\equiv -E=g_{\m\n}k^\m_t U^\n,
\la{EKNq}
\ee
\be
p_\vf\equiv L=g_{\m\n} k^\m_\vf U^\n.
\la{lphiKNq}
\ee
The geodesic four-velocity $U^\m\equiv \fr{dx^\m}{d\l}$ with $\l$ being the proper-time parameter satisfies the normalization 
\be
g_{\m\n} \fr{dx^\m}{d \l} \fr{dx^\n}{d \l}=-s^2 
\la{massshell}
\ee
with $s=1,0$ for the time-like and light-like cases, respectively. The geodesic four-velocity was obtained in the EF coordinates long ago in \cite{Carter:1968rr}. 
Denoting $U^\m=(\dot{u},\dot{z}, \dot{\th},\dot{\f})$ where the dot represents $\fr{d}{d\l}$,
the four-velocity components are given by 
\be
\dot{u}=\fr1{\fr1{z^2}+a^2\cos^2\th}\Big[-a(L+a E \sin^2\th)+\left(\fr1{z^2}+a^2\right)\D_z^{-1}(P+\sqrt{R} ) \Big],
\la{SigmaDotuz}
\ee
\be
\dot{z}=\fr{z^2}{\fr1{z^2}+a^2\cos^2\th}\sqrt{R} \,,
\la{zdotEF}
\ee
\be
 \dot{\th}=\pm\fr1{\fr1{z^2}+a^2\cos^2\th}\sqrt{\Th}
\la{SigmaDotthz}
\ee
\be
 \dot{\vf} 
=\fr1{\fr1{z^2}+a^2\cos^2\th}\Big[\left(aE+\fr{L}{\sin^2\th}\right)-a\D_z^{-1}(P+\sqrt{R})\Big],
\la{SigmaDotphiz}
\ee
with
\be
\D_z=a^2+z^{-2}-\fr{2M}{z},
\la{triangl}
\ee
\be
P=-aL -(a^2+z^{-2})E ,
\la{Pdef}
\ee
\be
\Th=\mathcal{K}-(L+Ea)^2-\cos^2 \th \left[a^2(s^2-E^2)+\fr{L^2}{\sin^2\th} \right],
\la{Thdef}
\ee
\be
R=P^2-\D_z \left(\mathcal{K}+\fr{s^2}{z^2} \right),
\la{Rdef}
\ee
where $E$, $L$ are the energy and the angular momentum of a particle of the mass $\m$, respectively; $\mathcal{K}$, called the Carter constant, is another integral of motion. Since a free-falling observer moves towards the black hole, we have chosen the negative brach.

In computing \rf{Qadef} one also needs the expressions for the covariant velocities (or momenta):
\be
U_u=-E,\quad U_z =\fr1{z^2 \D_z}\left(P(z)+\sqrt{R} \right),\quad
U_\th = \pm\sqrt{\Th},\quad U_\f=L.
\la{pzKN}
\ee
Let us also obtain the geodesic in the BL coordinates. One way to do this is to use the coordinate transformation:
\be
U^{BL}_\a =\fr{\pa x_{EF}^\b}{\pa x_{BL}^\a} U^{EF}_\b ,
\ee
where we put the subscripts and superscripts $EF,\,BL$ for clarity. On account of \rf{tphi2uvarphii}, which we quote below,
\be
du=dt+\fr{\Xi}{\D_r} (r^2+a^2) dr,\qquad d\f=-d\vf-\fr{\Xi a}{\D_r} dr
\la{tphi2uvarphiiq}
\ee
one gets, by substiting the BL components \rf{pzKN} into right-hand side,
\be
U^{BL}_r 
=-\fr{\sqrt{R}}{\D_r}.
\ee
Therefore the pole structure of $U^{BL}_r$ is the same as $\sim \fr{1}{\D_r}$. Repeating the steps one can show
\bea
U^{BL}_t &=& -E,\quad
U^{BL}_\th = \pm \sqrt{\Th},\quad  
U^{BL}_\vf = -L .
\eea
Near $z_{EH}$, $U_\m$ scales as $\sim \fr1{\k^2}$, which comes from the $\fr1{\D_r}$ factor of $U_r$. These results are confirmed by the direct BL-coordinate analysis presented in Appendix C.

\section{Quantum gravitational radiation}

Finally, we are ready to evaluate, at the quantum level, the momentum density eq. \rf{Qadef}, which we quote below for convenience: 
\be
P_\m=-{h_\m}^\r T_{\r\s}\, U^\s. 
\la{Qadefq}
\ee
The four-velocity $u^\m$ in \rf{Qadef} has been replaced by the geodesic four-velocity $U^\m$ associated with the observer. We evaluate $P_\m$ by taking the $Q=0,\L_0$ system for simplicity. The qualitative features of the results thus obtained should remain the same for more general cases.

As in our previous works, only the classical part of the one-loop stress tensor \rf{quanset} is needed for the computation to the leading order of the quantum corrections. (As for the solution, the quantum-corrected solution must be substituted.) This is due to the fact noted in section 2 that the classical matter parts of the quantum solution vanish. 
Explicitly, the classical form of the stress-energy tensor is
\bea
&& \hspace{-.2in}{ T_{\m\n}^{(class)}} = - \fr2{\k^2}\L g_{\m\n}+g_{\m\n}\Big[-|\pa_\r \psi-iqA_\r \psi|^2 -{\l}\Big(|\psi|^2+\fr{1}{2\l} \n^2\Big)^2 -\fr14 F_{\r\s}^2   \Big] \nn\\
&&\hspace{.2in} +  \left[(\pa_\mu \psi-iqA_\m \psi)(\pa_\n \psi^*+iqA_\n \psi^*)+(\m \leftrightarrow \n)\right]+  F_{\m\r}F_\n{}^\r.
\eea 
Another useful structure of the momentum density \rf{Qadefq} is revealed by explicitly writing out ${h_\m}^\r$:
\be
P_\m=-(\d_\m^\r+U_\m U^\r) T_{\r\s} U^\s= -T_{\m\r} U^\r- U_\m (U^\r T_{\r\s}\, U^\s).
\la{Qadefexp}
\ee
As we will soon see, the leading behavior comes from the second term:
\be
P_\m\sim - U_\m (U^\r T_{\r\s}\, U^\s)=- U_\m\, \r ,
\la{Qadeflead}
\ee
where $\r\equiv T_{\m\n} \,U^\m U^\n$ is the energy density, which has been computed in our previous work, \cite{Nurmagambetov:2019mih}. Below we evaluate the momentum density in the BL coordinates:
\be
P^{BL}_\m\sim - U^{BL}_\m\, \r \,.
\la{QadefleadBL}
\ee

\vspace{.1in}

\subsection{Evaluation of the momentum density}

Let us review how the scaling of $\r$ was determined. The structure of the action \rf{emsactcasetwo} is such that the matter terms come at higher order of $\k^2$ in the metric field equation in  \rf{quanfe}. This implies that rescalings of the matter fields are necessary for correct $\k$-scalings of various physical quantities. Put another way, the values of the matter fields take `ordinary' numbers, i.e., numbers that are not too big or small, when specified in terms of the appropriately $\k$-rescaled dimensionless fields which we denoted by $\xi,a_m$:
\bea
\psi=\fr{\xi}{\k}\quad,\quad A_m=\fr{a_m}{\k}, \quad m=1,2,3\,.
\eea
The fields $\xi, a_m$ have the following series expansions, the $\k$-rescaled versions of \rf{vecYsersim}: 
\bea
\xi(u,z,\th)&=& \tilde{\xi}_0 +\k^2\Big[\tilde{\xi}_0^h(u,\th) +  \tilde{\xi}_1^h(u,\th) Y+ \tilde{\xi}_2^h(u,\th)Y^2+\tilde{\xi}_3^h(u,\th) Y^3 +\cdots \Big] ,  \nn\\
\hspace{-.5in} a_1(u,z,\theta)&=& 
\k^2  \Big[ \at_{z0}^h(u,\th) +\at_{z1}^h(u,\th) Y+ \at_{z2}^h(u,\th)Y^2+\at_{z3}^h(u,\th)Y^3 + ...\Big] ,\nn\\
a_2(u,z,\theta)&=& \k^2  \Big[ \at_{\th0}^h(u,\th) +\at_{\th1}^h(u,\th) Y+ \at_{\th2}^h(u,\th)Y^2+\at_{\th3}^h(u,\th)Y^3 + ...\Big] ,\nn\\
a_3(u,z,\theta)&=& \k^2  \Big[ \at_{\f0}^h(u,\th) +\at_{\f1}^h(u,\th) Y+ \at_{\f2}^h(u,\th)Y^2+\at_{\f3}^h(u,\th)Y^3 + ...\Big].\nn\\
\la{vecYsersimres}
\eea
More generally, we denote the modes of the $\Yh$-series (cf. \rf{zzeroser}) with hats:
\bea
\xi(u,z,\th)&=& \hat{\xi}_0 +\k^2\Big[\hat{\xi}_0^h(u,\th) +  \hat{\xi}_1^h(u,\th) \Yh+ \hat{\xi}_2^h(u,\th)\Yh^2+\hat{\xi}_3^h(u,\th) \Yh^3 +\cdots \Big] ,  \nn\\
\hspace{-.5in} a_1(u,z,\theta)&=& 
\k^2  \Big[ \ah_{z0}^h(u,\th) +\ah_{z1}^h(u,\th) \Yh+ \ah_{z2}^h(u,\th)\Yh^2+\ah_{z3}^h(u,\th)\Yh^3 + ...\Big] ,\nn\\
a_2(u,z,\theta)&=& \k^2  \Big[ \ah_{\th0}^h(u,\th) +\ah_{\th1}^h(u,\th) \Yh+ \ah_{\th2}^h(u,\th)\Yh^2+\ah_{\th3}^h(u,\th)\Yh^3 + ...\Big] ,\nn\\
a_3(u,z,\theta)&=& \k^2  \Big[ \ah_{\f0}^h(u,\th) +\ah_{\f1}^h(u,\th) \Yh+ \ah_{\f2}^h(u,\th)\Yh^2+\ah_{\f3}^h(u,\th)\Yh^3 + ...\Big].\nn\\
\la{Ysershat}
\eea
It turns out that the leading modes $(\tilde{\xi}^h_0,\at_{z0}^h, \at_{\th0}^h, \at_{\f0}^h)$ play an important role in the energy. Since the location of the horizon at the quantum level, $z_{EH}^q$, (whose precise determination will be pursued elsewhere) will take the form of
\bea
z_{EH}^q=z_{EH}+ {\cal O}(\k^{2}),
\eea 
where $z_{EH}$ denotes the classical location of the event horizon, $\dot{t}$ should scale as $\dot{t}\sim {\cal O}(\k^{-2})$ at $z=z_{EH}^q$. With this scaling one gets, for the leading behavior of $\r$,
\bea
T_{\m\n} \;U^\m U^\n \sim \fr{\k^2 f(\tilde{\xi}^h_0,\at_{z0}^h, \at_{\th0}^h, \at_{\f0}^h)}{\k^4} \sim \fr1{\k^2} \,,
\la{mainres}
\eea
where $f(\tilde{\xi}^h_0,\at_{z0}^h, \at_{\th0}^h, \at_{\f0}^h)$ is a quantity that is proportional to $T_{00}$, whose explicit expression is given by
\bea
{f(\tilde{\xi}^h_0,\at_{z0}^h, \at_{\th0}^h, \at_{\f0}^h)} &=& \frac1{\sin^2\theta  \left(a^2 z_{EH}^2 \cos^2\theta +1\right)} 
\Big({a^2 z_{EH}^6 \sin^4\theta  (\pa_u \tilde{a}^h_{z0})^2+z_{EH}^2 (\pa_u \tilde{a}^h_{\f0})^2} \nn\\
&&\hspace{-1.3in} +\sin^2\theta  \Big[z_{EH}^4 (\pa_u \tilde{a}^h_{z0})^2 \left(a^2 z_{EH}^2 \cos^2\theta -2 M {z_{EH}}+1\right)   +2 \pa_u \tilde{\xi}^{h*}_0 \pa_u \tilde{\xi}^h_0 \left(a^2 z_{EH}^2 \cos^2\theta +1\right) \nn\\
	&&\hspace{1in}+2 a z_{EH}^4 \pa_u \tilde{a}^h_{z0} \pa_u \tilde{a}^h_{\f0}+z_{EH}^2 (\pa_u \tilde{a}^h_{\th0})^2\Big] \Big) +\cdots.  \la{feq}
\eea
The ellipses represent the terms higher in $\k$ or $\hbar$.

\vspace{.1in}

With the reminder above, let us now turn to the evaluation of $P^{BL}_\m$. We examine two aspects: its $\k$-scaling and $\th$-dependence. As we have pointed out in section 3, the leading behavior of $U_\m$ is
\be
U_\m \sim \fr1{\k^2} \,.
\ee
Note that this come from the $\m=r$, i.e., the radial direction. Combining with the result of $\r$, one gets, for the $\k$-scaling of $P^{BL}_\m$, 
\be
P^{BL}_\m \sim \fr1{\k^4} \,.
\ee 
The leading behavior of each term in \rf{Tabsplitg} is determined by the four-velocity and is the same as $\fr1{\k^6}$. The momentum density $P_\m$ contains a factor of a four-velocity, and this is why it has the higher inverse $\k$-scaling order, compared with that of the energy density $\r$.

For the $\th$-dependence, note that $P_\m$ is to be integrated over a two-sphere to eventually compute the flux; thus the proper quantity to look at is $P^{BL}_r \sin^2\th$. Also, since $U_r$ -- which gives the leading $\k$-scaling among $U_\m$ components -- does not have $\th$-dependence, one can simply examine the $\th$-dependence of $\r$ (therefore the disk-like and jet structures are already manifest in $\r$), which in turn is determined by $f(\tilde{\xi}^h_0,\at_{z0}^h, \at_{\th0}^h, \at_{\f0}^h)$:
\bea
P^{BL}_r \sin^2\th \sim f\sin^2\th .
\eea
Before getting to the algebra, let us develop some physical pictures, namely along the astrophysical context. The time evolution of the system will depend on various parameters, including the boundary conditions. Since the series solution technique is expected to work well with small deformations, one has to see to it that the function $f$ is examined in the corresponding regime.
The variables of $f$ are $\th,a,M$, and the quantum modes. Therefore one should consider a regime in which, roughly speaking, the mode values are small compared with $a$ and $M$. This will ensure that the perturbation preserves the underlying Kerr geometry. One should thus carry out a large $a$ and/or $M$ expansion of $f\sin^2\th$. The regime of the parameter space becomes narrower once we consider an astrophysical black hole: they are expected to rotate very fast, $a\sim M$. Another simplifying (but natural) condition is weak $\th$-dependence of the modes, $(\tilde{\xi}^h_0,\at_{z0}^h, \at_{\th0}^h, \at_{\f0}^h)$. They play the role of boundary modes, and their strong $\th$-dependence would, undesirably, make things weigh toward the perturbation instead of the black hole.

With this primer, we examine $f$ in the regime specified by the following two conditions. Firstly, we consider weak $\th$-dependence of the quantum deformations, except the $\th\ll 1$ or $|\fr{\pi}2-\th|\ll 1$ or $\pi-\th\ll 1$ region where more care is needed when taking various limits. We consider the region where $\th \centernot\ll 1$ and $|\fr{\pi}2-\th|\centernot\ll 1 $ and $\pi-\th \centernot\ll 1$. The weak $\th$-dependence condition is a simple and convenient setup to study the subsequent evolution of the original Kerr black hole. Secondly, we consider a fast-rotating limit with a large black hole mass of the original Kerr. In other words,
we consider $a\sim M\gg 1$. The large mass condition is to ensure that the black hole deformations are small. The fast rotation condition is a natural one since most of the astronomical black holes must rotate fast. In this regime (see \rf{feq}), 
\bea
\hspace{-.2in} f \sin^2\th \;\sim \;
2 \sin^2\th |\pa_u\tilde{\xi}^h_0|^2
+\frac{\sin^2\th\, z_{EH}^2 (\pa_u\at^h_{\th0})^2+z_{EH}^2 (\pa_u\at^h_{\f0})^2}{ a^2 (1-\sin^2\th) z_{EH}^2+1}+\cdots  \la{fse}
\eea
and thus $\pa_\th f$ takes
\be
\pa_\th (f \sin^2\th)\;\sim\;  \sin2\th \Big[ 2  |\pa_u\tilde{\xi}^h_0|^2
+\frac{ (z_{EH}^2+a^2 z_{EH}^4) (\pa_u\at^h_{\th0})^2+a^2z_{EH}^4 (\pa_u\at^h_{\f0})^2}{ (a^2 [1-\sin^2\th] z_{EH}^2+1)^2}\Big]+\cdots
\ee
where we have not explicitly recorded the terms on which $\pa_\th$ acts on the matter modes.
The square bracket term will be dominated by the first term and the extrema are determined by the $\sin2\th$-factor: the minimum occurs at $\th=0,\pi$ and the maximum at $\th=\fr{\pi}{2}$. This indicates that for the extrema one should consider the region $\th\ll 1$ or $|\fr{\pi}2-\th|\ll 1$ or $\pi-\th\ll 1$. However, these extrema are overshadowed by the divergent behaviors at $\th=0,\fr{\pi}2,\pi$, as we will now see. Let us turn to the region $\th\ll 1$ or $|\fr{\pi}2-\th|\ll 1$ or $\pi-\th\ll 1$. Unlike the region just considered, it is not obvious whether or not the terms on which $\pa_\th$ acts on the matter modes can be disregarded. However, this doesn't matter since $f\sin^2\th$ clearly shows a blowing-up behavior, and it is not necessary to examine its $\th$-derivative. We will focus on the small $\th$; the story for the $\th\sim \fr{\pi}2$ or $\th\sim \pi$ region is similar. When the angle $\th$ is not too small (or too close to $\fr{\pi}2,\pi$), one can take the modes as small as desired and the corresponding value of $f \sin^2\th$ will remain bounded. In contrast, in $\th\ll 1$ region, it is not possible to keep the modes bounded, and overall, the $f\sin^2\th$ value will blow up. To see this, it is useful to consider an arbitrary fixed location:
\be
z_0: \mbox{arbitrary fixed location} 
\ee
such that
\be
 z_0 \ll z_{EH} \;\; \mbox{or}\;\; r_0=\fr1{z_0}\gg r_{EH}.
\ee
Although we obtain the expression $f$ in \rf{feq} for $z_{EH}$, the corresponding expression for $z_0$ is given by the same function $f$ (put differently, \rf{feq} was obtained by substituting $z_0=z_{EH}$ in the following),
\bea
{f(\hat{\xi}^h_0,\ah_{z0}^h, \ah_{\th0}^h, \ah_{\f0}^h)} &=& \frac1{\sin^2\theta  \left(a^2 z_{0}^2 \cos^2\theta +1\right)} 
\Big({a^2 z_{0}^6 \sin^4\theta  (\pa_u \hat{a}^h_{z0})^2+z_{0}^2 (\pa_u \hat{a}^h_{\f0})^2} \nn\\
&&\hspace{-1.3in} +\sin^2\theta  \Big[z_{0}^4 (\pa_u \hat{a}^h_{z0})^2 \left(a^2 z_{0}^2 \cos^2\theta -2 M {z_{0}}+1\right)   +2 \pa_u \hat{\xi}^{h*}_0 \pa_u \hat{\xi}^h_0 \left(a^2 z_{0}^2 \cos^2\theta +1\right) \nn\\
	&&\hspace{.5in}+2 a z_{0}^4 \pa_u \hat{a}^h_{z0} \pa_u \hat{a}^h_{\f0}+z_{0}^2 (\pa_u \hat{a}^h_{\th0})^2\Big] \Big) +\cdots.  \la{feqzzero}
\eea
In the $M\gg 1, a\sim M$ regime, one gets
\bea
\hspace{-.2in} f \sin^2\th \;\sim \;
2 \sin^2\th |\pa_u\hat{\xi}^h_0|^2
+\frac{z_{0}^2\Big[\sin^2\th\,  (\pa_u\ah^h_{\th0})^2+ (\pa_u\ah^h_{\f0})^2\Big]}{ a^2 (1-\sin^2\th) z_{0}^2+1}+\cdots \la{fsezzero}
\eea
which is nothing but \rf{fse} with with $z_{EH}$ replaced by $z_0$. What is important is the fact that eventually things will be dominated by $\th \ra0$ limit. Let us focus on the first term of \rf{fsezzero} since the second term is subleading. (Inclusion of the second term does not change the qualitative conclusion.) We now reason that the first term, and therefore $f \sin^2\th$ as a whole, increases without bound as one approaches the small-$\th$ region. The crux of the argument is that the modes appearing in the $\Yh$ series in \rf{Ysershat}, in particular, $\hat{\xi}^h_0$, have such a characteristic. The series \rf{Ysershat} can still be made to converge by sufficiently narrowing the range of $z$ around $z_0$. In other words, one can just look at things right around $z_0$. To see that $\hat{\xi}^h_0$ blows up in the small-$\th$ region, let us recall the feature noted in section 2.1: various inverse powers of $\sin\th$ appear in some of the modes. The modes results in \rf{moderesh} have been obtained by setting $z_0=0$, the asymptotic boundary. By comparing the series expansions of $\xi$ for $z_0=0$ and $z_0\neq 0$ cases one can easily deduce that the hatted modes of $\xi$ field are given by linear combinations of the unhatted modes: $\hat{\xi}^h_0$, among other fields, is expressed as a linear combination of the entire tower of $\xi_i^h$:
\be
\hat{\xi}^h_0=\hat{\xi}^h_0(\xi_i^h), \; i=0,1,2,...
\ee
As noted in section 2.1, some of $\xi_i^h$ contain inverse powers of $\sin\th$, and thus imply divergence of $\hat{\xi}^h_0$.

To summarize, the momentum vector will have a controlled behavior outside of the region $\th\ll 1$ or $|\fr{\pi}2-\th|\ll 1$ or $\pi-\th\ll 1$. However, the solution displays strong emissions near the poles and equator. To our knowledge, these findings are not inconsistent with actual observations. Further out, we believe that they have the potential to provides links to complete mechanisms for AGN jets and accretion disks. The findings are qualitatively compared with observations and theoretical models below.

\vspace{.1in}
\subsection{Implications for AGN physics}

As well known in the astrophysical literature, AGNs have several characteristic features: enormous luminosity, accretion disk, and jets. It is also known that part of the UHECR, $\sim 10^9$ GeV, originate from AGNs. Although there has been significant progress (see, e.g., \cite{Abramowicz:2011xu}), their production apparatuses are not well understood currently. The common missing link in the mechanisms in these phenomena appears to be how extreme high energy particles are generated by the black hole. Since our result naturally leads to a bulk production of trans-Planckian radiation in the vicinity of the horizon, it may well provide the missing link. The bipolar emission certainly resembles jets of an AGN, and according to our solution the jets start at the event horizon. It may be possible to compare this with near-future x-ray observations.
As for the accretion disk observation, we are not aware of observational results that have probed the far inner structure of an accretion disk. Ideally, this would require use of an x-ray (or still better, $\g$-ray) probe. As for the theoretical side, our disk structure is different from the existing accretion disk theories in that ours displays strong radiation through the equator. Some of these issues are further discussed in the Conclusion.

\section{Conclusion}

Although it has been widely believed that the quantum gravitational effects are far too small and thus negligible, it has been explicitly demonstrated that that is not always the case \cite{Park:2017dib}\cite{Nurmagambetov:2018het}. Continuing our endeavour of exploring the near-horizon physics of a quantum-corrected solution of an Einstein-Maxwell-scalar system, in this work we have analyzed the momentum density vector. As in our recent sequels, we have analyzed the one-loop quantum-gravitational corrections. To demonstrate the robustness of the method employed, we have started by generalizing our previous results in several directions: we have extended the analysis to the previously uncovered cases of the non-vanishing cosmological constant and black hole charge. As the main task we have analyzed the quantum-gravitational Poynting vector and examined its near horizon behaviors. The momentum density scales as $\fr1{\k^4}$ and displays the structures that are in line with an accretion disk and jets of an AGN.

As noted in \cite{Nurmagambetov:2019mih}, time-dependence is crucial for the trans-Planckian scaling.
Then there is the question of how generically the time-dependence occurs. As commented on in section 2, the quantum-deformed solution can be time-independent. However, the time-independent quantum-deformed solutions will be of measure-zero compared to the time-dependent ones.

\vspace{.2in}
There are several future directions.

\vspace{.1in}

The fact that the $z$-expansion works well whereas the $Y$-expansion does not may be a reflection of the fact that the physical states are determined by the boundary degrees of freedom. Not unrelated to this, it seems tempting to relate the metric boundary mode(s) to the field of the reduced Lagrangian obtained in \cite{Park:2018xtt}. It will be interesting to make this potential connection more accurate.

A relatively urgent direction is to study the quantum effects on the apparent horizon, event horizon, and singularity. It should be possible, at least for some simpler systems, to explicitly determine the AH and EH. With the trans-Planckian energy at the EH, one will not have access to the singularity. Nevertheless, it will be of some interest to examine the singularity at the quantum level. In particular, our setup may provide an arena to tackle the question of whether or not the quantum effects resolve the singularity. For this purpose, it will be necessary to consider \rf{Ysershat} with $z_0$ approaching infinity. More work will be required to see whether or not one could make things sensible in such a limit.


Further exploration of the implications of our result to AGN physics will be worth it. The picture that we have for the accretion disks and jets is as follows. An astrophysical accretion disk of a black hole will play the role of sending the matter passing down to the innermost circular orbit. It is then the quantum gravitational process that produces the trans-Planckian energy and radiation from the infalling matter. 

There are several things to be further investigated to make the connection more plausible. For instance the energy scale of UHECRs, $\sim 10^9$ GeV, is lower than the Planck scale by a factor of $10^{10}$. The lion's share of this difference should be attributed to the potential energy loss over the black hole potential. Also, in the astrophysical literature there are results that seem consistent with a corona structure. It will be of some interest to study whether certain types of coronas may be a byproduct of the influence of the jets on the complex environs of a real astrophysical black hole. Still another potentially interesting issue is the fact that in the case of the Einstein-scalar system whose modes relations are listed in Appendix A, inverse powers of $\cos\th$ do not appear. This implies that the disk structure of the system is much weaker than that of the Einstein-Maxwell-scalar system. It is not clear at this point whether or not this absence of $\sec\th$ factors is due to the lower orders of the analysis. It may well be. If it is not an artifact of the low orders, it seems to be at odds with the astrophysical literature where the magnetic field plays an important role in jet formation.

\newpage
\appendix

\renewcommand{\theequation}{A.\arabic{equation}}
\setcounter{equation}{0}

\section{4D perturbation (Einstein-scalar case with CC)}

In our previous works we only considered the axisymmetric deformations. In section 2.2, extension to the full 4D deformations is discussed. For this part of the exercise, we take the following Einstein-scalar system:
\begin{equation}
\begin{array}{lll}
&S&=\fr1{\k^2}\int d^4x \sqrt{-g}\;\Big[R-2\L\Big] +\int d^4 x \sqrt{-g}\;  \Big[c_1  R^2+c_2 R_{\m\n}R^{\m\n} +\cdots\Big] \\[1ex]
&&  -\int d^4 x \sqrt{-g}\;\Big[|\pa_\mu \psi|^2 
+{\l}\Big(|\psi|^2+\fr{1}{2\l} \n^2\Big)^2   \Big].
\end{array}
\end{equation}
By taking the ansatz in eqs. \rf{dz}-\rf{zetaser4D}, one gets, after going through the `usual routines,'
\bea
&&\psi _0(u,\theta,\phi)=\psi _0,\quad \psi^*_0(u,\theta,\phi)=-\frac{\nu ^2}{2 \lambda  \psi _0},\quad \psi _1(u,\theta,\phi)=0\nn\\
&&\psi _2(u,\theta,\phi)=0,\quad \psi _3(u,\theta,\phi)=0\nn\\
&& \Delta _{-4}(u,\theta,\phi)=-\frac{\Lambda_0}{3},\quad \Delta _{-3}(u,\theta,\phi)=0,\quad
\Delta _{-2}(u,\theta,\phi)=1\nn\\
&&\Delta _{-1}(u,\theta,\phi)=0,\quad\Delta _0(u,\theta,\phi)=0,\quad \Delta _1(u,\theta,\phi)=0\nn\\
&&\Phi _{-1}(u,\theta,\phi)=1,\quad\Phi _0(u,\theta,\phi)=0,\quad  \Phi _1(u,\theta,\phi)=0\nn\\
&&\Phi _2(u,\theta,\phi)=0,\quad \Phi _3(u,\theta,\phi)=0,\quad \Phi _4(u,\theta,\phi)=0
\nn\\
\eea

\bea
&& \psi^{h*}_0(u,\theta,\phi)=\frac{\nu ^2 \psi^h_0(u,\theta ,\phi )+{ 4 c_3 \Lambda_0 \psi _0}}{2 \lambda  \psi _0^2},\quad  \nn\\
&&\psi^{h*}_1(u,\theta,\phi)=\frac{2 \Lambda_0 \psi^h_1(u,\theta ,\phi )+3 \nu ^2 \psi^h_1(u,\theta ,\phi )+6 \pa_u\psi^h_0(u,\theta ,\phi )}{6 \lambda  \psi _0^2}\nn\\
&&\psi^{h*}_2(u,\theta,\phi)=\frac1{6 \lambda  \psi _0^2}\Big(\left[2 \Lambda_0+3 \nu ^2\right] \psi^h_2(u,\theta ,\phi )\nn\\
&&\hspace{1in} +3 \left[\pa_\th^2\psi^h_0(u,\theta ,\phi )+\cot \theta\;  \pa_\th\psi^h_0(u,\theta ,\phi )+\csc ^2\theta \; \pa_\f^2\psi^h_0(u,\theta ,\phi )\right] \Big)\nn\\
&&\psi^{h*}_3(u,\theta,\phi)=\frac1{2 \lambda  \psi _0^3}(\nu ^2 \psi _0 \psi^h_3(u,\theta ,\phi )+\psi _0\, \pa_\th^2\psi^h_1(u,\theta ,\phi )-2 \psi _0\, \pa_u \psi^h_2(u,\theta ,\phi )\nn\\
&&\hspace{1in}+\psi _0 \cot \theta\;  \pa_\th\psi^h_1(u,\theta ,\phi )+\psi _0 \csc ^2\theta \; \pa_\f^2\psi^h_1(u,\theta ,\phi ))\nn\\
&&\Delta^h_{-4}(u,\theta,\phi)=\frac{1}{3} \left(-2 \Lambda_0 \Phi^h_{-1}(u,\theta ,\phi )- { \Lambda_1}\right),\quad \Delta^h_{-3}(u,\theta,\phi)=2 \pa_u\Phi^h_{-1}(u,\theta ,\phi )\nn\\
&&\pa_u \Delta^h_{-2}(u,\theta,\phi)=2 \pa_u\Phi^h_{-1}(u,\theta ,\phi ) \nn\\
&& \pa_u\pa_\f \Phi^h_{-1}(u,\theta,\phi)=0,\quad\pa_u\pa_\th\Phi^h_{-1}(u,\theta,\phi)=-2 \cot \theta\;  \pa_u\Phi^h_{-1}(u,\theta ,\phi )\nn\\
&&\Phi^h_0(u,\theta,\phi)=-\frac{3 \pa_u\Phi^h_{-1}(u,\theta ,\phi )}{\Lambda_0},\quad
 \pa_u \Phi^h_{1}(u,\theta,\phi)=0\nn\\
&&\Phi^h_1(u,\theta,\phi)=-\frac3{4 \Lambda_0}(\Delta^h_{-2}(u,\theta ,\phi )-\pa_\th^2\Phi^h_{-1}(u,\theta ,\phi )-3 \cot \theta\;  \pa_\th\Phi^h_{-1}(u,\theta ,\phi )\nn\\
&&\hspace{2in}+\csc ^2\theta  \pa_\f^2\Phi^h_{-1}(u,\theta ,\phi )) \nn\\
&&\pa_\th\Phi^h_2(u,\theta,\phi)=-2 \cot \theta \; \Phi^h_2(u,\theta ,\phi )\nn\\
&&\Phi^h_3(u,\theta,\phi)=-\frac3{4 \lambda  \Lambda_0^2 \psi _0^2}(2 \lambda  \Lambda_0 \psi _0^2 \Delta^h_0(u,\theta ,\phi )+3 \lambda  \psi _0^2 \Delta^h_{-2}(u,\theta ,\phi )\nn\\
&&\hspace{.3in}-3 \lambda  \psi _0^2 \,\pa_\th^2\Phi^h_{-1}(u,\theta ,\phi )-9 \lambda  \psi _0^2 \cot \theta \; \pa_\th\Phi^h_{-1}(u,\theta ,\phi )+3 \lambda  \psi _0^2 \csc ^2\theta  \;\pa_\f^2\Phi^h_{-1}(u,\theta ,\phi ))\nn\\
&& \pa_\f\Phi^h_4(u,\theta,\phi)=0.
\eea

\newpage

\renewcommand{\theequation}{B.\arabic{equation}}
\setcounter{equation}{0}

\section{Splitting of each sector of stress tensor}

The 3+1 splitting of the stress tensor is given in eqs. \rf{Tabsplitg}. Although the form of the momentum density in \rf{Qadef} is sufficient for the further evaluation in section 4, splitting of each sector provides additional insights, and may be useful for future research. The splitting of the Maxwell's sector has been reviewed in section 3.1. Here we carry out the splitting of the scalar and one-loop graviton sectors.

Note that the stress tensor \rf{quanset} is such that the classical part consists of matter terms whereas the quantum correction is purely of the graviton sector. Denoting the classical part by $T^{(class)}$ and quantum part by $T^{(quan)}$, it can be written as 
\be
{
T_{\m\n}=T^{(class)}_{\m\n}+ T^{(quan)}_{\m\n},
}
\la{Tcorr}
\ee
where 
{
\bea
T^{(class)}_{\m\n}&=&  - \fr2{\k^2}\L g_{\m\n}+g_{\m\n}\Big[-|\pa_\r \psi-iqA_\r \psi|^2 -{\l}\Big(|\psi|^2+\fr{1}{2\l} \n^2\Big)^2 -\fr14 F_{\r\s}^2  \Big] \nn\\
&+&\Big[ { \left((\pa_\mu \psi-iqA_\m \psi)(\pa_\n \psi^*+iqA_\n \psi^*)+(\m \leftrightarrow \n)\right)}+  F_{\m\r}F_\n{}^\r  \Big],
\eea
}
\be
\hspace{-1.5in} { T^{(quan)}_{\m\n} }=  g_{\m\n} \Big(c_1R^2-(4c_1+c_2)\nabla^2 R  +c_2 R_{\r\s}R^{\r\s}\Big) \la{tquan}
\ee 
\[    
 \hspace{.6in}   \;\;-2 \Big(2c_1 RR_{\m\n}  -(2c_1+c_2) \nabla_\m \nabla_\n R
-2c_2 R_{\k_1\m\n\k_2} R^{\k_1\k_2}+c_2\nabla^2 R_{\m\n}\Big)  +\cdots.   
\]
One can also split the classical part: 
\[
{
T^{(class)}_{\m\n}\equiv T^{(scalar)}_{\m\n}+T^{(e.m.)}_{\m\n},
}
\]
where
\[
{
T^{(e.m.)}_{\m\n} \equiv F_{\m\r}{F_\n}^\r-\fr14 g_{\m\n} F_{\r\s}F^{\r\s},
}
\] 
{
\bea
T^{(scalar)}_{\m\n} \equiv \left[{D}_\m \y \bar{D}_\n \y^*+{D}_\n \y \bar{D}_\m \y^* \right]- g_{\m\n}\left[{D}_\r\y \bar{D}^\r\y^*+\l\left(|\y|^2+\fr1{2\l} \n^2\right)^2 \right]  \nn\\
\eea
}
with
\be
{
{D}_\m\y\equiv \pa_\m \y-iq A_\m \y,\qquad \bar{D}_\m \y^*\equiv \pa_\m \y^*+iq A_\m \y^*\,.
}
\la{calDdef}
\ee

\subsubsection*{scalar sector}

By applying \rf{Tabsplitg} to the scalar sector, one gets
\be
{
T_{\m\n}^{(scalar)}=\left(T_{\r\s}^{(scalar)} u^\r u^\s \right) u_\m u_\n+\left(P_\m^{(scalar)} u_\n+u_\m P_\n^{(scalar)} \right)+{h_\m}^\r{h_\n}^\s T_{\r\s}^{(scalar)}.
}
\la{Tabsplitscalar}
\ee
By working out the three terms in \rf{Tabsplitscalar}, one gets
{
\[
T_{\m\n}^{(scalar)}=u_\m u_\n \Big[2{\cal D}\y\, \bar{\cal D} \y^*+{D}_\r\y \bar{ D}^\r\y^*+\l\big(|\y|^2+\fr1{2\l} \n^2\big)^2\Big]
\]
\[
-2u_{(\m} \bar{\cal D}_{\n)}\y^* \,{\cal D} \y-2 u_{(\m}{\cal D}_{\n)} \y\,\bar{\cal D} \y^*
\]
\be
-h_{\m\n}\Big[{D}_\r\y \bar{D}^\r\y^*+\l\Big(|\y|^2+\fr1{2\l} \n^2\Big)^2\Big]+{\cal D}_\m \y \bar{\cal D}_\n \y^*+{\cal D}_\n \y \bar{\cal D}_\m \y^*\,
\la{Tscalarsplit}
\ee
}
where
\be
{
{\cal D} \equiv u^\r {D}_\r,\,\,\,\bar{\cal D}\equiv u^\r \bar {D}_\r,\,\,\, {\cal D_\m}\equiv {h_\m}^\n {D}_\n,\,\,\, \bar{\cal D}_\m\equiv {h_\m}^\n \bar{D}_\n.
}
\ee
From \rf{Tscalarsplit} one reads off
\be
{
P^{(scalar)}_\m  =- \bar{\cal D}_{\m}\y^*\, {\cal D} \y
- {\cal D}_{\m} \y\,\bar{\cal D} \y^*.
}
\ee

\subsubsection*{one-loop graviton sector}

For the graviton sector \rf{tquan}, we split, for convenience, {$T^{(quan)}_{\m\n}$} as
\be
{
T^{(quan)}_{\m\n}=g_{\m\n}\,T_1-T_{2\,\m\n},
}
\la{Thab1}
\ee
\be
{
T_1=c_1 R^2-(4c_1+c_2)\nabla^2 R+c_2 R_{\r\s}R^{\r\s},
}
\la{Th1}
\ee
\be
{
T_{2\,\m\n}=2\left(2c_1 R R_{\m\n}-(2c_1+c_2)\nabla_\m \nabla_\n R-2c_2 R_{\r\m\n\s}R^{\r\s}+c_2 \nabla^2 R_{\m\n} \right).
}
\la{Th2ab}
\ee
Further noting that
\be
{
g_{\m\n}\,T_1=h_{\m\n} T_1-u_\m u_\n T_1,
}
\la{gabT1h}
\ee
and
\be
{
T_{2\,\m\n}={h_\m}^{\r} {h_\n}^{\s} T_{2\,\r\s}+2P^{(q)}_{(\m} u_{\n)}+u_\m u_\n \,\r^{(q)},
}
\la{Th2absplit}
\ee
where
\be
{
\r^{(q)}\equiv T_{2\,\r\s} \,u^\r u^\s 
}
\la{Th2tildef}
\ee
\be
{
P^{(q)}_\m=-{h_\m}^\r\, T_{2\,\r\s} u^\s .
}
\la{Phadef}
\ee
With this, the quantum part of the stress tensor $T^{(q)}_{\a\b}$ can be rewritten
	\be
	{
	T^{(quan)}_{\m\n}=-\left(\r^{(g)}+T_1 \right)u_\m u_\n-2P^{(q)}_{(\m} u_{\n)}+T_1h_{\m\n}
	-{h_\m}^{\r} {h_\n}^{\s} T_{2\,\r\s}.
	}
	\la{Thabsplit}
	\ee
The Riemann tensor splits as follows \cite{Costa:2012cw},
\be
{
{R^{\m\n}}_{\r\s}=4{{\cal E}^{[\m}}_{[\r}u_{\s]}u^{\n]}+2\Big({\e^{\t\l}}_{\r\s}u_\l {\cal H}_{\t}^{[\n}u^{\m]} +\e^{\t\m\n\l} u_\l {\cal H}_{\t[\s}u_{\r]}\Big)
+\e^{\m\n\vf\w}u_\w {\e^{\t\l}}_{\r\s}{\cal F}_{\vf\t}u_\l ,
}
\la{Riemannirrep}
\ee
where
\be
{
{\cal E}_{\r\s}\equiv R_{\r\m\s\n} u^\m u^\n,
}
\la{Eabdef}
\ee
\be
{
{\cal H}_{\r\s}\equiv \ast R_{\r\m\s\n}u^\m u^\n=\fr12 {\e_{\r\m}}^{\e\t}R_{\e\t\s\n} u^\m u^\n,
}
\la{Habdef}
\ee
\be
{
{\cal F}_{\r\s}\equiv \ast R \ast_{\r\m\s\n}u^\m u^\n=\fr14 {\e^{\m\n}}_{\r\e}{\e^{\l\t}}_{\s\w} R_{\m\n\l\t} u^\e u^\w.
}
\la{Fabdef}
\ee
It follows from these that {${\cal E}_{\r\s}={\cal E}_{\s\r}$, ${\cal F}_{\r\s}={\cal F}_{\s\r}$; ${\cal H}_{\r\s}$ is of arbitrary symmetry, but trace-free: ${{\cal H}_{\s}}^\s=0$.}
Taking the trace over the 1st and the 3rd indices gives the Ricci tensor:
\be
{
{R^\r}_\s={\cal E}_\m^\m\, u^\r u_\s-{{\cal E}^\r}_\s-\e^{\t\r\m\n}{\cal H}_{\m\t}u_\s u_\n-\e_{\t\s\m\n}{\cal H}^{\m\t}u^\r u^\n-{{\cal F}^\r}_\s+{\cal F}_\m^\m \,{h^\r}_\s.
}
\la{Ricciirrep}
\ee
For the Ricci scalar, one gets
\be
R=2({\cal F}^\s_\s-{\cal E}^\s_\s)  \equiv 2(\Tr{\cal F}-\Tr{\cal E}).
\la{Rirrep}
\ee
We consider two examples. The first example is {$R_{\m\n}R^{\m\n}$ that appears in $T_1$}; one can show, after some algebra, 
\be
{ R_{\m\n}R^{\m\n}}=\left(\Tr {\cal E}\right)^2+\left(\Tr {\cal F} \right)^2-2\Tr {\cal F} \,\Tr {\cal E}+\Tr({\cal E}^2)+2\Tr({\cal E}{\cal F})+\Tr({\cal F}^2)+4\Tr \tilde{{\cal H}}^2,
\la{TrRR1}
\ee
where { $\tilde{{\cal H}}_{\r\s}$ is the antisymmetric part of ${\cal H}_{\r\s}$ and $\Tr \tilde{\cal{H}}^2=\tilde{\cal H}^{\r\s}\tilde{\cal H}_{\s\r}$}. For the second example, let us consider {$\nabla_\m\nabla_\n R$ that appears in $T_{2\,\m\n}$}: defining
\be
{
\tilde{\nabla}_\m \equiv {h_\m}^{\m'}\nabla_{\m'}, \qquad \tilde{\nabla} \equiv u^{\m'}\nabla_{\m'} 
}
\la{tildenabdef}
\ee
it can be expressed as
{
\bea
\nabla_\m\nabla_\n R &=& u_\m u_\n \tilde{\nabla}^2 R+\Big[\tilde{\nabla}_\m \tilde{\nabla}_\n R-(\tilde{\nabla}_\m u_\n)\tilde{\nabla} R\Big] \nn\\
&&\hspace{-.2in}-\Big[u_\m \tilde{\nabla} \tilde{\nabla}_\n R+u_\n \tilde{\nabla}_\m \tilde{\nabla} R-u_\m (\tilde{\nabla} u_\n) \tilde{\nabla} R\Big].
\la{nabnabR}
\eea
}

\newpage

\renewcommand{\theequation}{C.\arabic{equation}}
\setcounter{equation}{0}

\section{Kerr spacetime geodesics in the BL}

Because we couldn't find a review in the literature that contains all of the BL geodesic results used in section 4, we review them here. The metric we consider is the pure Kerr metric with $l^{-2}=0$ and $Q=0$ in \rf{ds2KNBL}. 
The metric admits two Killing vectors:
\be
k^\m_t=(1,0,0,0),\qquad k^\m_\vf=(0,0,0,1),
\la{KillingKN}
\ee
which leads to two integrals to the geodesic equations: the energy
\be
p_t\equiv -E=g_{\m\n}k^\m_t U^\n,
\la{EKN}
\ee
and angular momentum projection
\be
p_\vf\equiv L_{BL}=g_{\m\n} k^\m_\vf U^\n ,
\la{lphiKN}
\ee
where we have introduced $L_{BL}$ to distinguish it from $L$ of the EF; due to the sign conventions of the azimuthal angles, one has $L_{BL}=-L$. Following Carter \cite{Carter:1968rr}, the rest of the momenta (recall, $p_\m=g_{\m\n} \dot x^\n$, where the dot denotes the derivative with respect to the proper-time $\l$) come from the Hamilton-Jacobi (HJ) equation 
\be
\fr{\pa S}{\pa \l}=H\left(p=\fr{\pa S}{\pa x},x \right)=\fr12 g^{\m\n} \left(\fr{\pa S}{\pa x^\m} \right)\left(\fr{\pa S}{\pa x^\n} \right)
\la{HJeq}
\ee
with $H(p,x)=1/2 g^{\m\n}p_\m p_\n=-\fr12 s^2$ ($s=0,1$ are the null and time-like geodesics, respectively) and with 
\be
S=-\fr12 s^2 \l-E t+L_{BL} \vf+S_\th(\th)+S_r(r).
\la{SBL}
\ee
By substituting \rf{SBL} into the HJ equation \rf{HJeq} one gets
\be
g^{tt}\left(\fr{\pa S}{\pa t}\right)^2+2 g^{t\vf} \fr{\pa S}{\pa t}\fr{\pa S}{\pa \vf}+g^{\vf\vf}\left(\fr{\pa S}{\pa \vf} \right)^2+g^{rr} \left(\fr{\pa S}{\pa r}\right)^2+g^{\th\th}\left(\fr{\pa S}{\pa \th}\right)^2+s^2=0.
\la{HJKerrBL}
\ee
Upon substituting the explicit form of the inverse metric, eq. \rf{HJKerrBL} turns into
\[
\D_r \left(\fr{\pa S_r}{\pa r}\right)^2+ \left(\fr{\pa S_\th}{\pa \th}\right)^2-\left[\fr{(r^2+a^2)^2}{\D_r}-a^2 \sin^2\th \right] E^2+\fr{4Mr a}{\D_r}\,EL_{BL}
\]
\be
+\left(\fr1{\sin^2\th}-\fr{a^2}{\D_r} \right) L_{BL}^2+s^2 (r^2+a^2 \cos^2\th)=0.
\la{HJKerrBL3}
\ee
By rewriting
\be
\fr{4Mr a}{\D_r}\,EL_{BL}=-2aEL_{BL}+2aEL_{BL}\,\fr{r^2+a^2}{\D_r},
\la{idKerr BL}
\ee
the $\th$-dependent part of \rf{HJKerrBL3} becomes
\be
\left(\fr{\pa S_\th}{\pa \th}\right)^2+a^2 s^2 \cos^2\th+\left(a\sin\th E-\fr{L_{BL}}{\sin\th} \right)^2={\cal K}
\la{StheqBL}
\ee
with the Carter constant ${\cal K}$, which is a separation constant. Since $p_\th=\pa S_{\th}/\pa \th$, eq. \rf{StheqBL} yields
\be
\quad p_\th=\pm \sqrt{\Th},
\la{pthBL}
\ee
where
\be
\Th = {\cal K}-(Ea-L_{BL})^2-\cos^2\th \left[a^2(s^2-E^2)+\fr{L_{BL}^2}{\sin^2\th} \right].
\la{ThdefBL}
\ee
The $r$-dependent part of \rf{HJKerrBL3} gives
\be
\D_r \left(\fr{\pa S_r}{\pa r}\right)^2+s^2 r^2-\fr1{\D_r} \left[(r^2+a^2)E-aL_{BL}\right]^2=-{\cal K},
\la{SreqBL}
\ee
which can be written as
\be
\left(\fr{\pa S_r}{\pa r}\right)^2=\fr1{\D^2_r}\left[P^2(r)-\D_r({\cal K}+s^2 r^2) \right]
\la{SreqBL1}
\ee
where 
\be
P(r)\equiv -(r^2+a^2)E+aL_{BL}.
\ee
Noting $p_r=\pa S_r/\pa r$ and introducing a new variable $R(r)$,
\be
R(r)=P^2(r)-\D_r({\cal K}+s^2 r^2)\; 
\la{RdefBL}
\ee
one gets
\be
p_r=\pm\fr{\sqrt{R}}{\D_r}.
\la{prBL}
\ee
To sum up, the geodesic equation in the BL is integrated to yield
\be
p_t=U_t=-E,\quad p_r=U_r=\pm \fr{\sqrt{R}}{\D_r},\quad  p_\th=U_\th=\pm\sqrt{\Th},\quad p_\vf=U_\vf=L_{BL}.
\la{psKerrBLr}
\ee
As indicated above, the $p_r$ (or the covariant velocity $U_r$) has two branches. Since a free-falling observer moves towards the black hole, the negative branch $U_r=-\sqrt{R}/\D_r$ has been chosen in sections 3 and 4.

\newpage

\end{document}